%% file: Hybridcostafin.tex
\pdfoutput=1
\documentclass[journal,final,letterpaper,oneside,onecolumn,11pt]{IEEEtran}
\usepackage{latexsym,url}
\usepackage{amsmath,amssymb,graphicx}% ,makeidx}
\usepackage{citesort}
\usepackage{epsf}

\input{macros}

\renewcommand{\baselinestretch}{1.0}

\begin{document}
\title{Joint Source Channel Coding with Side Information Using
Hybrid Digital Analog Codes}
\author{
 Makesh Pravin Wilson \\
Dept. of Electrical Engineering \\
Texas A\&M University \\
College Station, TX 77843 \\
{\tt makesh@ece.tamu.edu} \and\\
 Krishna Narayanan \\
Dept. of Electrical Engineering \\
Texas A\&M University \\
College Station, TX 77843 \\
{\tt krn@ece.tamu.edu} \and\\
 Giuseppe Caire \\
 Dept. of Electrical Engineering \\
University of Southern California \\
Los Angeles, CA 90089\\
{\tt caire@usc.edu} \thanks{This work was supported by the
National Science Foundation under grant CCR 0515296.} } \maketitle

\begin{abstract}

We study the joint source channel coding problem of transmitting
an analog source over a Gaussian channel in two cases - (i) the
presence of interference known only to the transmitter and (ii) in
the presence of side information known only to the receiver. We
introduce hybrid digital analog forms of the Costa and Wyner-Ziv
coding schemes. Our schemes are based on random coding arguments
and are different from the nested lattice schemes by Kochman and
Zamir that use dithered quantization. We also discuss superimposed
digital and analog schemes for the above problems which show that
there are infinitely many schemes for achieving the optimal
distortion for these problems. This provides an extension of the
schemes by Bross et al to the interference/side information case.
We then discuss applications of the hybrid digital analog schemes
for transmitting under a channel signal-to-noise ratio mismatch
and for broadcasting a Gaussian source with bandwidth compression.

\end{abstract}

%%% ----------------------------------------------------------------------
\maketitle
%%% ----------------------------------------------------------------------
\section{Introduction and Problem Statement}

Consider the classical problem of transmitting $K$ samples of a
discrete-time independent identically distributed (i.i.d) real
Gaussian source $\vv$ in $N$ uses of an additive white Gaussian
noise (AWGN) channel such that the mean-squared error distortion is
minimized. Let the source be encoded into the sequence $\xv$ which
satisfies a power constraint $E[\xv \xv^T] \leq NP$. Let us first
consider the case of $K=N$ and let the output of the AWGN channel
$\yv$ be given by
\[
\yv = \xv + \wv
\]
where $\wv$ is a noise vector of i.i.d Gaussian random variables
with zero mean and variance $\sigma^2$. If the source variance is
$\sigma_v^2$, then the optimal mean-squared error distortion that
can be achieved is $D_{opt} =
\frac{\sigma_v^2}{1+\frac{P}{\sigma^2}}$. This optimal performance
can be achieved by two very simple schemes. The first one is
separate source and channel coding, where the source is first
quantized and the quantization index is transmitted using an
optimal code for the AWGN channel. The second scheme is uncoded
(analog) transmission with power scaling
\cite{goblick65,gastpar03}, where the source is not explicitly
quantized. Recently, it was shown by Bross, Lapidoth and Tinguely
\cite{shraga06} that there is a family of infinitely many schemes
that are optimal, which contains the separation based scheme and
uncoded (analog) transmission as special cases.

In this paper, we consider the problem of transmitting $K$ samples
of an i.i.d Gaussian source through $N$ uses of an AWGN channel. We will refer to
the ratio of $N/K$ as the bandwidth efficiency $\lambda$.
We first consider the case of $\lambda = 1$ and the
presence of an interference known only to the transmitter and/or
side information available only at the receiver.
We derive hybrid
digital analog (HDA) coding schemes for these cases where the source
is not explicitly quantized and show that they can obtain the
optimal distortion. These can be viewed as the equivalent of uncoded
transmission but in the presence of an interference or side
information. Then, we show that there is a family of infinitely many
schemes that are optimal for this problem which contain pure
separation based schemes and HDA schemes as special cases. This can
be viewed as the extension of Bross, Lapidoth and Tinguely's
\cite{shraga06} result in the presence of
interference/side-information. An interesting aspect of the hybrid
digital analog coding schemes proposed here is that they do not
require binning unlike their separation based counterparts.

The HDA scheme proposed here for the case of interference known at
the transmitter is closely related to the scheme considered by
Kochman and Zamir in \cite{kochman06}, although this was developed
independently. The difference is that the proposed scheme is based only on random
coding arguments and does not use nested lattices like in
\cite{kochman06}. As a result, the relationship between the
auxiliary random variable and the source is made more explicit. We
also consider several applications of the HDA schemes which are not
considered in \cite{kochman06}. We consider the non-asymptotic SNR
case unlike in \cite{kochman06}. Further, the performance of this
scheme in the presence of SNR mismatch is analyzed. Finally, the use
of a HDA Costa based scheme for broadcasting a Gaussian source to
two users with bandwidth compression, where $\lambda < 1$ is discussed.
In the case of side-information available only at the receiver, the
proposed scheme is similar to the scheme in \cite{reznic05} and
again uses random coding arguments instead of nested lattices.

The paper is organized as follows. First in
Section~\ref{sec:jsccinterference}, we discuss the problem of
transmitting an i.i.d Gaussian source in the presence of a
Gaussian interference known only to the transmitter. We introduce
a hybrid digital analog (HDA) Costa coding scheme where the source
is not explicitly quantized and show that this is optimal. We then
discuss a generalized HDA Costa coding scheme and show that there
infinitely many schemes that are optimal.  In
Section~\ref{sec:jsccwynerziv}, we discuss similar schemes and
results for the case of having side information available only at
the receiver (Wyner-Ziv problem) and in
Section~\ref{sec:combined}, briefly consider the situation having
interference known only to the transmitter and side-information
available only at the receiver. In \cite{merhav03}, Merhav and
Shamai have shown that separate Wyner-Ziv coding followed by
Gelfand-Pinsker coding is optimal for this problem. However, we
show that there is a joint source-channel coding scheme for the
case of Gaussian source, interference and side information. This
result in Section~\ref{sec:combined} is a fairly straightforward
extension of the results in Section~\ref{sec:jsccinterference} and
Section~\ref{sec:jsccwynerziv}, but is included for completeness
and to make the exposition clear. In
Section~\ref{sec:analysisSNRmismatch}, we study the performance of
these schemes when the SNR of the channel is different from the
designed SNR and in Section~\ref{sec:distexp} the distortion
exponents of these schemes are analyzed. In
Section~\ref{sec:gcbc}, we consider the problem of transmitting a
Gaussian source in the absence of an interference, but when the
channel bandwidth is smaller than the source bandwidth and show
how the HDA Costa coding scheme is useful. Finally, in
Section~\ref{sec:broadcastbandwidthcomp}, we consider the problem
of broadcasting a Gaussian source to two users through AWGN
channels and propose a joint source channel coding scheme based on
HDA Costa coding.

We use the following notation in this paper. Vectors are denoted
by bold face letters such as $\xv$. Upper case letters are used to
denote scalar random variables. When considering a sequence of
i.i.d random variables, a single upper case letter is used to
denote each component of the random vector.

\section{Transmission of a Gaussian source over a Gaussian channel with interference known
only at the transmitter} \label{sec:jsccinterference}

\begin{figure}[htbp]
\begin{center}
\includegraphics[scale=0.5,angle =-90]{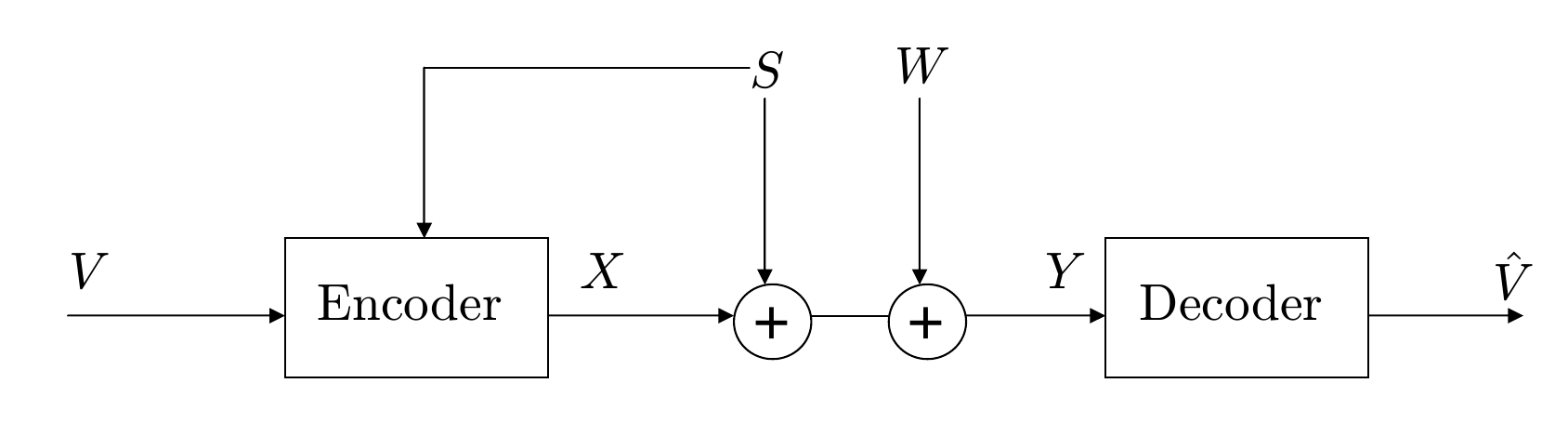}
\end{center}
\caption{Block diagram of the joint source channel coding problem
with interference known only at the transmitter.}
\label{fig:jscccostaproblem}
\end{figure}

We first consider the problem of transmitting $N$ samples of a
real analog source $\mathbf{v}
 \in
\mathbb{R}^N$ (this corresponds to $K=N$)
, with components $V$ which are independent Gaussian
random variables with $V \sim {\mathcal {N}}(0,\sigma_v^2)$ in $N$
uses of an AWGN channel with noise variance $\sigma^2$ in the
presence of an interference $\mathbf{s} \in \mathbb{R}^N$ which is
known to the transmitter but unknown to the receiver. Further, let
us assume that $S$'s are a sequence of real i.i.d Gaussian random
variables with zero mean and variance $Q$ and let the input power
to the channel $\EE[X^2]$ be constrained to be $P$. The problem
setup is shown schematically in Fig.~\ref{fig:jscccostaproblem}.
The received signal $\yv$ is given by
\begin{equation}
\yv = \xv + \sv + \wv
\end{equation}
where $\sv$ is the interference and $\wv$ is the AWGN.

The optimal distortion of
$\frac{\sigma_v^2}{\left(1+\frac{P}{\sigma^2}\right)}$ can be
obtained even in the presence of the interference by using the
following (obvious) separate source and channel coding scheme.

\subsection{Separation based scheme with Costa coding (Digital Costa Coding)}
We first quantize the source using an optimal quantizer to produce
an index $m \in \{1,2,\ldots,2^{NR} \}$, where $R = \frac 1 2 \log
\left(1+\frac{P}{\sigma^2} \right) - \epsilon$. Then, the index is
transmitted using Costa's writing on dirty paper coding scheme
\cite{costa}. Since the quantizer output is digital information,
we refer to this scheme as digital Costa coding. We briefly review
this here to make it easier to describe our proposed techniques
later on.

Let $U$ be an auxiliary random variable given by
\begin{equation}
\label{eqn:auxiliaryrvdca} U = X + \alpha S
\end{equation}
where $X \sim {\mathcal {N}}(0,P)$ is independent of $S$ and $\alpha
= \frac{P}{P+\sigma^2}$.

We first create an $N$-length i.i.d Gaussian code book
${\mathcal{U}}$ with $2^{N (I(U;Y) - \delta)}$ codewords, where each
component of the codeword is Gaussian with zero mean and variance
$P+\alpha^2 Q$. Then evenly (randomly) distribute these over
$2^{NR}$ bins. For each $\mathbf{u}$, let $i(\mathbf{u})$ be the
index of the bin containing $\mathbf{u}$. For a given $m$, we look
for an $\mathbf{u}$ such that $i(\mathbf{u}) = m$ and
$(\mathbf{u},\mathbf{s})$ are jointly typical. Then, we transmit
$\mathbf{x} = \mathbf{u} - \alpha \mathbf{s}$. Note that since
$(\uv,\sv)$ are jointly typical, from (\ref{eqn:auxiliaryrvdca}), we
can see that $\mathbf{x} \bot \mathbf{s}$ and satisfies the power
constraint.

The received sequence $\mathbf{y}$ is given by
\begin{equation}
\yv = \xv + \sv + \mathbf{w}
\end{equation}

At the decoder, we look for a $\uv$ that is jointly typical with
$\yv$ and declare $i(\uv)$ to be the decoded message. Since $R =
\frac 1 2 \log \left(1+\frac{P}{\sigma^2} \right) - \epsilon$, the
distortion in $\vv$ given by $D(R)$, where $D$ is the distortion
rate function. For a Gaussian source and mean squared error
distortion $D(R) = \sigma_v^2 2^{-2R}$ and, hence, the overall
distortion can be made to be arbitrarily close to
$\frac{\sigma_v^2}{\left(1+\frac{P}{\sigma^2}\right)}$ by a proper
choice of $\epsilon$ and $\delta$.

While the above scheme is straightforward, in the following three
sections we show that there are a few other joint source channel
coding schemes, which are also optimal. In fact, there are
infinitely many schemes which are optimal. Although, these schemes
are all optimal when the channel SNR is known at the transmitter,
their performance is in general different when there is an SNR
mismatch. The joint source channel coding schemes to be discussed
in the next sections have advantages over the separation based
scheme discussed in such a situation.

\subsection{Hybrid Digital Analog Costa Coding}
\label{subsec:hdacosta}
 Let us now describe a joint
source-channel coding scheme where the source $\vv$ is not
explicitly quantized. We refer to this scheme as hybrid digital
analog (HDA) Costa coding for which the code construction, encoding
and decoding procedures are as follows.

We first define an auxiliary random variable $U$ given by
\begin{equation}
\label{eqn:auxiliaryrvac} U = X + \alpha S + \kappa V
\end{equation}
where $X \sim {\mathcal {N}}(0,P)$ and $X$, $S$ and $V$ are pairwise
independent.

\begin{enumerate}

\item Codebook generation: Generate a random i.i.d code book
$\mathcal{U}$ with $2^{N R_1}$ sequences, where each component of
each codeword is Gaussian with zero mean and variance
$P+\alpha^2Q+\kappa^2 \sigma_v^2$.

\item Encoding: Given an $\sv$ and $\vv$, find a $\uv$ such that
$(\uv,\sv,\vv)$ are jointly typical with respect to the distribution
obtained from the model in (\ref{eqn:auxiliaryrvac}) and transmit
$\xv = \uv - \alpha \sv - \kappa \vv$. If such an $\uv$ cannot be
found, we declare an encoder failure. Let $P_{e_1}$ be the
probability of an encoder failure.

From standard arguments on typicality and its extensions to the
infinite alphabet case \cite{coverbook}, it follows that  $P_{e_1}
\rightarrow 0$ as $N \rightarrow \infty$ provided
\begin{eqnarray}
R_1 & > & I(U;S,V)\\
    &=& h(U) - h(U|S,V) \\
         &=& h(U) - h(X|S,V) \\
         &=& h(U) - h(X)\\
         \label{eqn:finalIUSVa}
         &=& \frac{1}{2} \log \frac{P + \alpha^2 Q + \kappa^2 \sigma_v^2}{P}
\end{eqnarray}
where the results follow because $X = U - \alpha S - \kappa V$ and
$X \bot S,V$. Notice that when a $\uv$ that is jointly typical
with $\sv$ and $\vv$ is found, $\xv$ satisfies the power
constraint.

\item Decoding : The received signal is $\yv = \xv + \sv + \wv$.
At the decoder, we look for an $\uv$ that is jointly typical with
$\yv$. If such a unique $\uv$ can be found, we declare $\uv$ as
the decoder output or, else, we declare a decoder failure. Let
$P_{e_2}$ be the probability of the event that the decoder output
is not equal to the encoded $\uv$ (this includes the probability
of decoder failure as well as the probability of a decoder error).

In order to analyze $P_{e_2}$, consider the equivalent communication
channel between $U$ and $Y$. Notice that we have in effect
transmitted a codeword $\uv$ from a random i.i.d codebook for $U$
with $2^{NR_1}$ codewords through the equivalent channel whose
output is $\yv$. Again, from the extension of joint typicality to
the infinite alphabet case, $P_{e_2} \rightarrow 0$ as $N
\rightarrow \infty$ provided that
\begin{eqnarray}
\nonumber
I(U;Y) & > & R_1 \\
\nonumber
I(U;Y) &=& h(U) - h(U|Y)\\
\nonumber
       &=& h(U) - h(U-\alpha Y |Y) \\
\nonumber
       &=& h(U) - h(X+ \alpha S + \kappa V - \alpha X - \alpha S - \alpha
       W | Y)\\
       \label{eqn:IUY}
       &=&h(U) - h(\kappa V + (1-\alpha) X - \alpha W | Y)
\end{eqnarray}
Now, let us choose
\begin{eqnarray}
\alpha & = & \frac{P}{P+\sigma^2} \\
\label{eqn:optimalalphakappa} \kappa^2 & = &
\frac{P^2}{(P+\sigma^2)\sigma_v^2} - \frac{\epsilon}{\sigma_v^2}
\end{eqnarray}
For the above choice of $\alpha$, it can be seen that
\[ \EE [
(\kappa V +(1-\alpha)X - \alpha W) Y] = 0 \] and, hence,
(\ref{eqn:IUY}) reduces to
\begin{eqnarray}
\nonumber
I(U;Y) &=& h(U) - h(\kappa V + (1-\alpha) X - \alpha W)\\
\label{eqn:finalIUY}
       &=& \frac{1}{2} \log \frac{P + \alpha^2 Q + \kappa^2\sigma_v^2}{P-\epsilon}
\end{eqnarray}

Hence, $P_{e_2}$ can be made arbitrarily small as long as
\begin{equation}
R_1 < \frac{1}{2} \log \frac{P + \alpha^2 Q +
\kappa^2\sigma_v^2}{P-\epsilon}
\end{equation}

Combining this with the condition for encoder failure, $P_{e_1}$ and
$P_{e_2}$ can both be made arbitrarily small provided
\begin{equation}
\frac{1}{2} \log \frac{P + \alpha^2 Q + \kappa^2 \sigma_v^2}{P} <
R_1 < \frac{1}{2} \log \frac{P + \alpha^2 Q +
\kappa^2\sigma_v^2}{P-\epsilon} \label{eqn:R1cond}
\end{equation}

Therefore, by choosing an $\epsilon_1$, $0 < \epsilon_1 <
\epsilon$ and $R_1 = \frac{1}{2} \log \frac{P + \alpha^2 Q +
\kappa^2\sigma_v^2}{P-\epsilon_1}$ we can satisfy
(\ref{eqn:R1cond}) and make $P_{e_1} \rightarrow 0$ and $P_{e_2}
\rightarrow 0$ as $N \rightarrow \infty$.

\item Estimation: If there is no decoding failure, we form the
final estimate of $\vv$ as an MMSE estimate of $\vv$ from $[\yv \
\uv]$. After some algebra this is given by ,

\begin{equation}
\hat{\vv} = \frac{\kappa \sigma_v^2}{P - \epsilon}(\uv - \alpha \yv)
\end{equation}

The distortion is then given by,

\begin{equation}
E[(V-\hat{V})^2] =\frac{\sigma_v^2}{1 + \frac{P}{\sigma^2}}
\frac{P}{P-\epsilon} \leq \frac{\sigma_v^2}{1 + \frac{P}{\sigma^2}}
+ \delta(\epsilon)
\end{equation}
with $\delta(\epsilon)$ is vanishing for arbitrarily small
$\epsilon$.
If an encoder or decoder failure was declared, we set the estimate
of $\vv$ to be the zero vector. However, as shown above the
probability of these events can be made arbitrarily small and,
hence, they do not contribute to the overall distortion, which can
be seen to be arbitrarily close to the optimal distortion achievable
in the absence of the interference.
\end{enumerate}

%\begin{thm}
%To the achieve the optimal distortion, the choice of $\alpha$ and
%$\kappa$ in (\ref{eqn:optimalalphakappa}) is unique.
%\end{thm}
%
%\underline{Proof:} Will be written later as it is not very important
%right now. This is still an interesting result because it appears
%that we can change $\alpha$ and $\kappa$ and obtain a region of
%achievable distortions with just one layer of the analog costa
%coding.

We have presented a joint source channel coding scheme in the
presence of an interference known only to the transmitter. The use
of the term hybrid digital analog Costa coding needs some
explanation. The scheme is not entirely analog in that the
auxiliary random variable is from a discrete codebook. However, in
contrast to digital Costa coding, the source is not explicitly
quantized and is embedded  into the transmitted signal $\xv$ in an
analog fashion. This is the reason for calling this as HDA Costa
coding and this has some interesting consequences which are
discussed in the following section.

Another feature of the HDA Costa coding scheme is that it does not
make use of binning, rather it needs a single quantizer codebook
that is also a good channel code. In practice, this may have some
impact on the design since there are ensembles of codes that are
provably good quantizers and channel codes \cite{litsyn05}. In the
Gaussian case, good lattices that are both good for coding and for
quantization are known. The binning approach however, requires a
nesting condition. That is, the fine code much be a good channel
code, but it much contain a subcode (or a coarse code) and its cosets that must be
good quantizers. This may be a  more difficult condition to obtain
in practice.

\subsection{Superimposed digital and HDA Costa coding scheme}
\label{subsec:superimp digi and hda costa}

\begin{figure}[htbp]
\begin{center}
\includegraphics[scale=0.4,angle = 0]{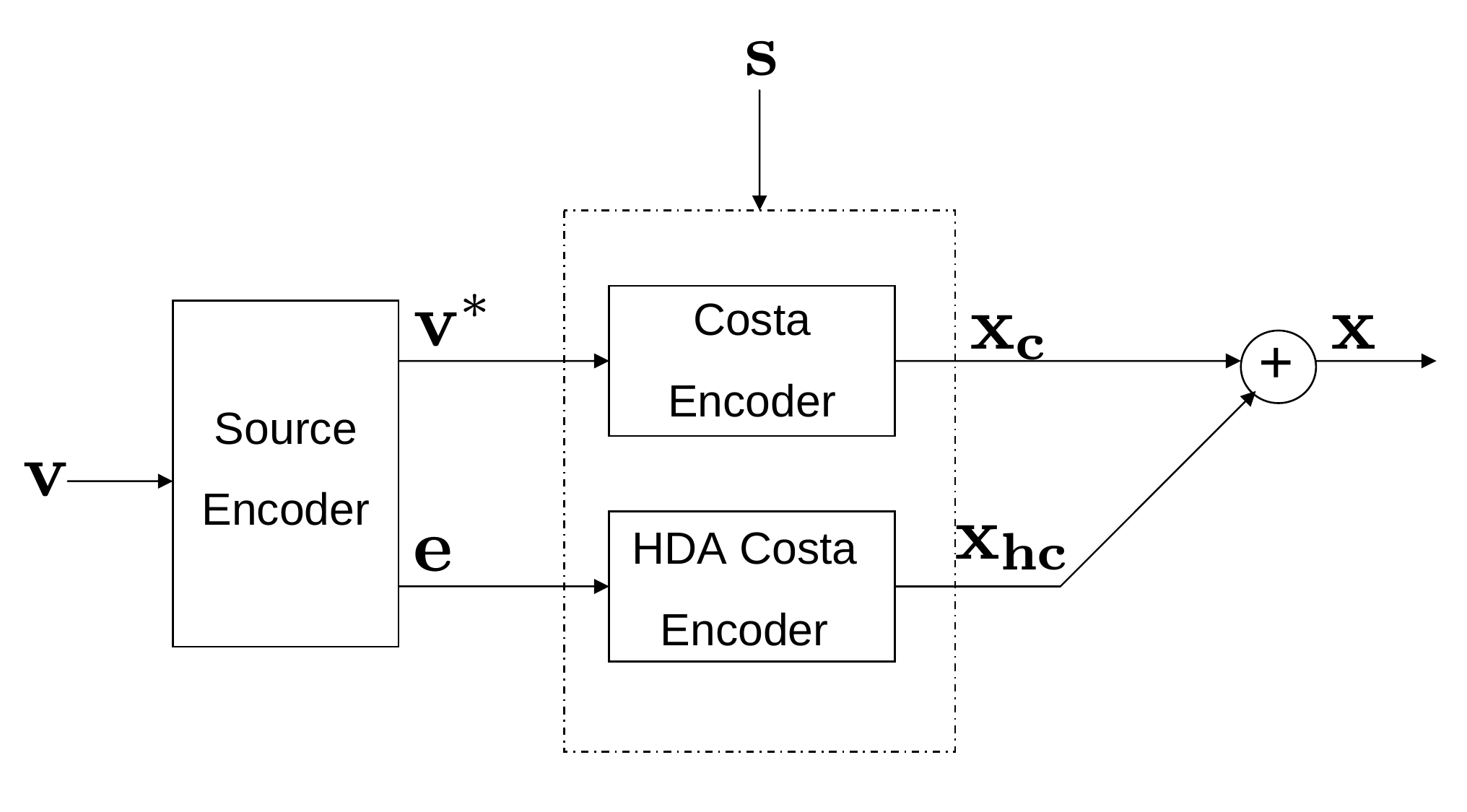}
\end{center}
\caption{Encoder model for superimposed coding}
\label{fig:supimp}\end{figure}

 Recently in \cite{shraga06},
Bross, Lapidoth and Tinguely considered the problem of transmitting
$N$ samples of a Gaussian source in $N$ uses of an AWGN channel, in
the absence of the interferer. They showed that there are infinitely
many superposition based schemes, which contain pure separation
based scheme and uncoded transmission as special cases. In this
section, we show that the same is true in the presence of an
interference also and show the corresponding scheme, which is given
in Fig.~\ref{fig:supimp}.
%The q index using digital costa. Then say abt U . Give big pic

The transmitted signal is a superposition of two signals
$\mathbf{x_c}$ and $\mathbf{x_{hc}}$, which are the outputs of a
digital Costa encoder and an HDA encoder, respectively.

The source is first quantized at a rate of $R < C$ using an optimal
source code and let the quantization error be $\ev = \vv-\vv^*$,
where $\vv^*$ is the reconstruction. The quantization error $\ev$
has a variance $\sigma_e^2 = \sigma_v^2 2^{-2R}$. The first stream
in Fig.~\ref{fig:supimp} is a digital Costa encoder that encodes the
quantization index by treating $\sv$ as interference and produces
the signal $\mathbf{x_c}$, which has a power of $P_C$. The second
stream is a HDA Costa encoder of rate $R$ which treats $\sv$ and $\mathbf{x_c}$
as interference and produces $\mathbf{x_{hc}}$, which has a power of
$P_{HC} = P-P_C$. The transmitted signal is the superposition (sum)
of $\mathbf{x_c}$ and $\mathbf{x_{hc}}$.

In the digital Costa encoder in the first stream, the auxiliary
random variable is given by $\mathbf{u_c} = \mathbf{x_c} + \alpha_c
\sv $ with $\mathbf{x_c}\bot \sv$.  A power of $P_C =
(P+\sigma^2)(1-2^{-2R})$ is used in the first stream and $\alpha_c $
is chosen as $\frac{P_C}{P_C + P_{HC} + \sigma^2}$. Note that this
corresponds to treating $\mathbf{x_{hc}}$ as noise in addition to
the channel noise.

In the second stream, the quantization error $\ev$ is encoded using
an HDA Costa coding scheme and a power of $P_{HC} = P - P_C =
(P+\sigma^2)2^{-2R} - \sigma^2$ is used. Note that since $R < C$,
the power $P_{HC}$ is always positive. The auxiliary random variable
is chosen as $\mathbf{u_{hc}} = \mathbf{x_{hc}} + \alpha_{hc}
(\mathbf{x_c} +\sv) +\kappa \mathbf{e} $, where $\mathbf{x_c} +\sv$
acts as the net interference. Hence, $\mathbf{x_{hc}}$ is chosen to
be independent of $\mathbf{x_c}$, $\sv$ and $\mathbf{e}$, and
$\alpha_{hc}$ is chosen to be $\frac{P_{HC}}{P_{HC}+\sigma^2}$.
$\kappa$ is chosen similar to (\ref{eqn:optimalalphakappa}) which
gives $\kappa^2 = \frac{P_{HC}^2}{(P_{HC}+\sigma^2)\sigma_v^2
2^{-2R}} - \frac{\epsilon}{\sigma_v^2 2^{-2R}}$.

At the decoder the quantization index from the first stream is first
decoded and the reconstruction $\mathbf{v^*}$ is obtained. Then, an
estimate of the quantization error $\ev$ is obtained from the second
stream using the HDA costa decoder. The overall distortion is the
distortion in estimating $\ev$. Using the analysis of the HDA Costa
scheme in Section~\ref{subsec:hdacosta}, this can be seen to be
\begin{equation}
D = \frac{\sigma_e^2}{1 + \frac{(P+\sigma^2)2^{-2R} -
\sigma^2}{\sigma^2}}+ \delta(\epsilon) = \frac{\sigma_v^2}{1 +
\frac{P}{\sigma^2}} + \delta(\epsilon)
\end{equation}

By choosing $\epsilon$ to be arbitrarily small we can make
$\delta(\epsilon) \rightarrow 0 $ and achievable a distortion of $D
= \frac{\sigma_v^2}{1 + \frac{P}{\sigma^2}} $, which is the optimal
distortion.

Note that for any source coding rate chosen in the first
stream namely $R$, the resulting distortion is optimal. By varying $R$, we
can get an infinite family of optimal joint source channel coding
schemes.

\subsection{Generalized Hybrid Costa coding}

In the previous section, we described a superposition technique. In
this section we show a scheme that does not explicitly do
superposition. Moreover this also introduces an interesting scheme
that is intermediate between HDA Costa having no bins to the digital
Costa having bins corresponding to the capacity of the channel.

Once again we quantize the source $\vv$ to $\mathbf{v^*}$ at a rate
$R$, that is strictly lesser than the channel capacity, using an
optimal vector quantizer. Let $\ev = \vv - \vv^*$ be the
quantization error vector. Note that for an optimal quantizer, as
the Rate-Distortion limit is approached,
the quantization error $\ev$ will be Gaussian.

We next define an auxiliary random variable $U$ given by
\begin{equation}
\label{eqn:auxiliaryrvhcc} U = X + \alpha S + \kappa_1 E
\end{equation}
where $X \sim {\mathcal {N}}(0,P), E \sim
{\mathcal{N}}(0,\sigma_v^22^{-2R})$, and $X$, $S$ and $E$ are
independent of each other. $\alpha$ and $\kappa_1$ are constants,
the choice of which is discussed below
\begin{enumerate}

\item Codebook generation: Generate a random i.i.d code book
$\mathcal{U}$ with $2^{N I(U;Y)}$ sequences, where each component of
each codeword is Gaussian with zero mean and variance
$P+\alpha^2Q+\kappa_1^2 \sigma_v^22^{-2R}$. These codewords are
uniformly distributed in $2^{NR}$ bins and this is shared between
the encoder and the decoder.

\item Encoding: Let $m$ be the quantization index corresponding to
the quantized source $\mathbf{v^*}$. Let $i(\mathbf{u})$ represent
the index of a bin that contains $\uv$. For a given $m$ find an
$\uv$ such that $i(\uv) = m$ and $(\uv,\sv,\ev)$ are jointly typical
with respect to the distribution in model
(\ref{eqn:auxiliaryrvhcc}). We next transmit the vector $\xv = \uv
-\alpha \sv - \kappa_1 \mathbf{e} $. Note that since $(\uv,\sv,\ev)$
are jointly typical, from (\ref{eqn:auxiliaryrvhcc}), we can see
that $\mathbf{x} \bot \mathbf{s},\mathbf{e}$ and satisfies the power
constraint.

\item Decoding : The received signal is $\yv = \xv + \sv + \wv$.
At the decoder, we look for an $\uv$ that is jointly typical with
$\yv$. If such a unique $\uv$ can be found, we declare $\uv$ as the
decoder output or, else, we declare a decoder failure. Next we make
an estimate of $\mathbf{e}$ from $\mathbf{u}$ and $\yv$.

We can see by similar Gelfand-Pinsker coding arguments that $R <
I(U;Y) - I(U;S,E)$. Note

\begin{eqnarray}
\label{eqn:finalIUSVb1} \nonumber
I(U;Y) & - & I(U;S,E) \\
\nonumber & = & h(U|S,E) - h(U|Y)\\
\nonumber    &=& h(X) - h(U-\alpha Y|Y) \\
\nonumber         &=& h(X) - h(\kappa_1 E + (1-\alpha) X - \alpha W  |Y) \\
\nonumber         & \stackrel{(a)}{=}& h(X) - h(\kappa_1 E + (1-\alpha) X - \alpha W)\\
         &=& \frac{1}{2} \log \left(\frac{P}{\kappa_1^2
         \sigma_v^22^{-2R} + (1-\alpha)^2P+\alpha^2 \sigma^2}\right) \\
         \label{eqn:finalIUSVb}
 \nonumber     & \stackrel{(b)}{>}& R
\end{eqnarray}

In (\ref{eqn:finalIUSVb1}) we choose $\alpha = \frac{P}{P+\sigma^2}
$ and $\kappa_1^2 =
\frac{P}{P+\sigma^2}\frac{(P+\sigma^2)-\sigma^22^{2R}}{\sigma_v^2} -
\frac{\epsilon ((P+\sigma^2)-\sigma^22^{2R})}{P \sigma_v^2} $. The
choice of $\alpha$ ensures $(1-\alpha)X - \alpha W$ is orthogonal to
$Y$ to get the equality in (a). $\kappa_1$ is chosen as above to
satisfy the inequality in (b).
 This shows that we can decode the codeword $\uv$
with a very high probability and we can decode the message $m =
i(\uv)$ and $\vv^*$.

\item Estimation: If there is no decoding failure, we form the
final estimate of $\vv$ as an MMSE estimate of $\vv$ from $[\mathbf{v^*} \ \uv \ \yv ]$. The estimate
can be obtained as follows. Let us define $\sigma_e^2 = \sigma_v^2 2^{-2R}$.
Let $\mathbf{\Lambda}$ be the covariance matrix of $[V^* \ U \
Y]^T$ and let $\mathbf{\Gamma}$ be the correlation vector between $V$ and
$[V^* \ U \ Y]^T$. Then, $\mathbf{\Lambda}$ and $\mathbf{\Gamma}$ are
given by

\[
 \mathbf{\Lambda} = \left(
                       \begin{array}{ccc}
                         \sigma_v^2-\sigma_e^2 & 0 & 0 \\
                         0 & P +  \kappa_1^2 \sigma_e^2 + \alpha^2 Q & P + \alpha Q  \\
                         0 & P + \alpha Q  & P + Q + \sigma
                         ^2 \\
                       \end{array}
                     \right) \ \mbox{and} \ \mathbf{\Gamma} = \left(
                                                   \begin{array}{ccc}
                                                     \sigma_v^2-\sigma_e^2 & \kappa_1 \sigma_e^2 & 0 \\
                                                   \end{array}
                                                 \right)^T
                                                 \].

The coefficients of the linear MMSE estimate are given by
$\mathbf{{\Lambda^{-1}}}\mathbf{\Gamma}$ and the minimum mean-squared error is given by
\[
D = \sigma_v^2 - {\mathbf{\Gamma}}^T \mathbf{{\Lambda^{-1}}}\mathbf{\Gamma} = \left(
\frac{\sigma_v^2}{1+\frac{P}{\sigma^2}}\right) +
\delta(\epsilon)
\]
where $\delta(\epsilon) \rightarrow 0$ as $\epsilon \rightarrow 0$. Thus, in the limit of
$\epsilon \rightarrow 0$, $D = \left(
\frac{\sigma_v^2}{1+\frac{P}{\sigma^2}}\right)$.
%The estimator is linear and the estimation
%error in $\mathbf{e}$ is $ \left(
%\frac{\sigma_v^2}{1+\frac{P}{\sigma^2}}\right) +
%\delta(\epsilon)$. Since $\vv = \mathbf{v^*} + \mathbf{e}$ and
%$\mathbf{v^*}$ is decoded correctly with a high probability the
%error in estimating $\vv$, $D = \left(
%\frac{\sigma_v^2}{1+\frac{P}{\sigma^2}}\right)+ \delta(\epsilon)$.
%Choosing
%an $\epsilon$ that is arbitrarily small $\delta(\epsilon)
%\rightarrow 0$ and $D = \left(
%\frac{\sigma_v^2}{1+\frac{P}{\sigma^2}}\right)$.
\\

\  \    It must be noted that this scheme is an intermediate
between digital Costa coding scheme with the maximum possible bins
equal to the capacity of the channel and the analog Costa coding
scheme with no bins. Thus we can get a family of schemes with
varying bins for the Gaussian channel.

The generalized hybrid Costa coding scheme appears to be closely
related to the superimposed digital and HDA Costa coding schemes.
The subtle difference however is in the generalized hybrid Costa
coding scheme, the transmitted signal $X$ is not a superposition
of two streams as seen in the superposition case.

%Another advantage of this is, if the source had been already
%quantized we can use an hybrid Costa code after the quantizer and
%still be optimal for the required SNR.
\end{enumerate}

\section{Transmission of a Gaussian source through a channel with
 side information available only at the receiver}\label{sec:jsccwynerziv}

 In this section we consider the problem of transmitting a discrete-time analog source
 over a Gaussian noise channel when the receiver has some side information
 about the source. This problem is a dual of the problem
 considered in the previous section and is considered here for the sake of
 completeness.  Consider the system model as follows. Let $\vv \in \mathbb{R}^N$ be the
discrete-time analog source where $V$'s are independent Gaussian
random variables with $V \sim {\mathcal {N}}(0,\sigma_v^2)$. Let
$\mathbf{v'} \in \mathbb{R}^N$ be the side information that is known
only at the receiver. The correlation between the source and the
side information is modeled as \begin{equation}
\label{eqn:auxiliaryrvdcc} V = V' + Z
\end{equation} where $Z \sim \mathcal{N}(0,\sigma_z^2)$ and $V'$ is i.i.d
Gaussian. Here $V'$ and $Z$ are mutually independent random
variables. The source $\vv$ must be encoded into $\xv$ and
transmitted over an AWGN channel and the received signal is
\begin{equation}
\yv = \xv +  \wv
\end{equation}
where $\xv$ satisfies a power constraint $P$ and $\wv$ is  AWGN
having a noise variance of $\sigma^2$. The following schemes can
be shown to be optimal for this case.

\subsection{Separation Based Scheme with Wyner Ziv Coding (Digital Wyner Ziv Coding)}
 One strategy is using a separation scheme with an optimal
 Wyner-Ziv code of rate $R$ followed by a channel code. We also refer to this scheme as
the digital Wyner-Ziv scheme. We briefly explain the digital
 Wyner-Ziv scheme and then establish our information theoretic model for the
 HDA Wyner-Ziv coding scheme.

\begin{figure}[htbp]
\begin{center}
\includegraphics[scale=0.5,angle=0]{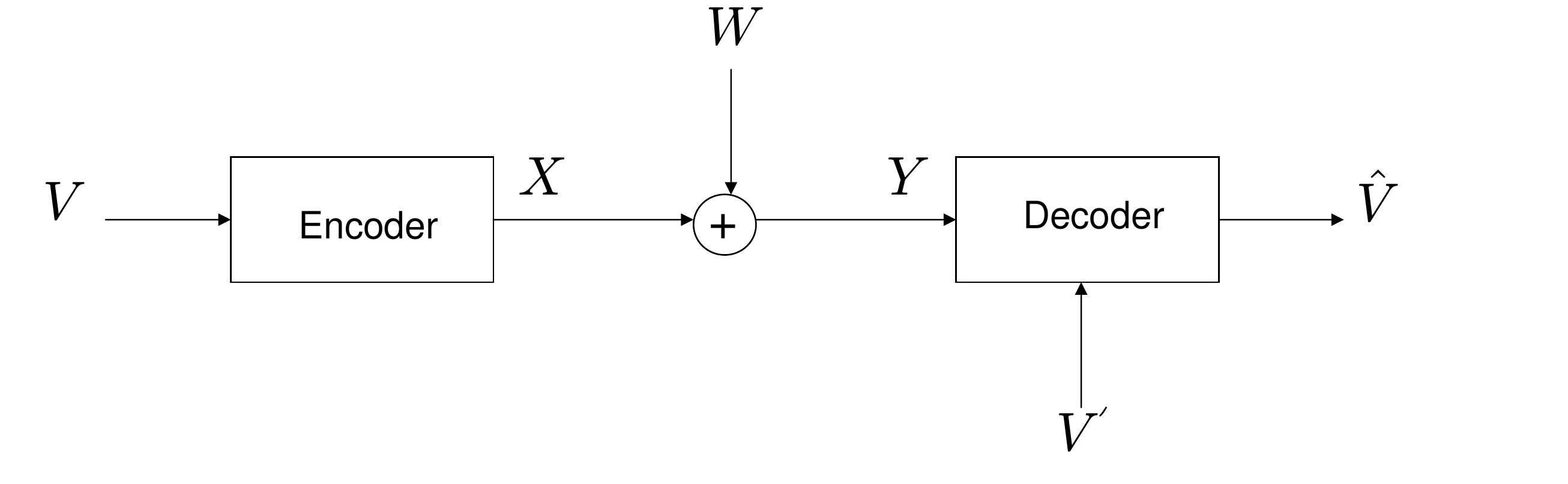}
\end{center}
\caption{Block diagram of the joint source channel coding problem
with side information known only at the receiver.}
\end{figure}

 Suppose
the side information is available both at the encoder as well as the
receiver, the best possible distortion is $D = \frac{\sigma_z^2}{1 +
\frac{P}{\sigma^2}}$. The same distortion can be achieved using the
following scheme and is a direct consequence of Wyner and Ziv's
result \cite{wyner74}. This can be achieved as follows,

Let $U$ be an auxiliary random variable given by
\begin{equation}
\label{eqn:auxiliaryrvdcb} U = \sqrt{\alpha} V + B
\end{equation} where $\alpha = 1 - \frac{D}{\sigma_z^2} = \frac{P}{P + \sigma^2}$ and $B \sim
\mathcal{N}(0,D)$. We create an $N$-length i.i.d Gaussian code book
${\mathcal{U}}$ with $ 2^{N I(U;V)}$ codewords, where each component
of the codeword is Gaussian with zero mean and variance $\alpha
\sigma_v^2 + D$ and evenly distribute them over $2^{NR}$ bins. Let
$i(\mathbf{u})$ be the index of the bin containing $\mathbf{u}$. For
each $\mathbf{v}$, find an $\mathbf{u}$  such that
$(\mathbf{u},\mathbf{v})$ are jointly typical. The index $
i(\mathbf{u})$ is the Wyner-Ziv source coded index. The index
$i(\mathbf{u})$ is encoded using an optimal channel code of rate
arbitrarily close to $\frac{1}{2} \log(1 + \frac{P}{\sigma^2})$ and
transmitted over the channel. At the receiver decoding of the index
$i(\mathbf{u})$ is possible with high probability as an optimal code
book for the channel is used. Next for the decoded $i(\mathbf{u})$
we look for an $\mathbf{u}$ in the bin whose index is
$i(\mathbf{u})$ such that $(\mathbf{u},\mathbf{v'})$ are jointly
typical. From $\mathbf{v'}$ and the decoded $\mathbf{u}$ we make an
estimate of the source $\vv$ as follows.

\begin{equation}
 \mathbf{\hat{v}} = \mathbf{v'} +  \sqrt{\alpha}(\uv - \sqrt{\alpha} \mathbf{v'})
\end{equation}

This yields the optimal distortion D.

\subsection{Hybrid Digital Analog Wyner Ziv Coding}
\label{subsec:hdawz} In this section, we discuss a different joint
source channel coding scheme that does not involve quantizing the
source explicitly. This scheme is quite similar to the modulo
lattice modulation scheme in \cite{reznic05}; the difference being
that a nested lattice is not used. The auxiliary random variable $U$
is generated as follows.

\begin{equation}
\label{eqn:auxiliaryrvdcd} U = X + \kappa V
\end{equation} where $\kappa$ is defined as  $\kappa^2 =
\frac{P^2}{(P+\sigma^2)\sigma_z^2} - \frac{\epsilon}{\sigma_z^2}$
and $X \sim \mathcal{N}(0,P)$.

\begin{enumerate}

\item Codebook generation: Generate a random i.i.d code book
$\mathcal{U}$ with $2^{N R_1}$ sequences, where each component of
each codeword is Gaussian with zero mean and variance $P+\kappa^2
\sigma_v^2$. This codebook is shared between the encoder and the
decoder.

\item Encoding: For a given $\vv$ find an $\uv$ such that
$(\uv,\vv)$ are jointly typical and transmit $\xv = \uv - \kappa
\vv$. This is possible with arbitrarily high probability if $R_1 >
I(U;V)$

\item Decoding: The received signal is $\yv = \xv + \wv$.
Find an $\uv$ such that $(\mathbf{v'},\yv,\uv)$ are jointly typical.
A unique such $\uv$ can be found with arbitrarily high probability
if $R_1 < I(U;V',Y)$. We next show below that we can choose an $R_1$
to satisfy $I(U;V) < R_1 < I(U;V',Y)$. This requires $I(U;V) <
I(U;V',Y)$ which can be shown as follows

\begin{eqnarray} \label{eqn:decodehdawz}
I(U;V',Y) & = & h(U) - h(U|V',Y) \nonumber\\
    &=& h(U) - h(U-\kappa V' - \alpha Y|V',Y) \nonumber \\
         &=& h(U) - h(\kappa Z + (1-\alpha) X - \alpha W  |V',Y) \nonumber \\
         &\stackrel{(a)}{=}& h(U) - h(\kappa Z + (1-\alpha) X - \alpha W) \nonumber \\
         &=& \frac{1}{2} \log \left(\frac{P + \kappa^2 \sigma_v^2}{\kappa^2
         \sigma_z^2 + (1-\alpha)^2P+\alpha^2\sigma^2}\right) \nonumber \\
         &\stackrel{(b)}{=}& \frac{1}{2} \log \left(\frac{P + \kappa^2\sigma_v^2}{P}\right) + \delta(\epsilon) \nonumber\\
         &=& h(U) - h(U|V) + \delta(\epsilon) \nonumber\\
         &=& I(U;V) + \delta(\epsilon)
\end{eqnarray}

In (\ref{eqn:decodehdawz}), (a) follows because $(\kappa Z +
(1-\alpha) X - \alpha W)$ is independent of $Y$ and $V'$. (b)
follows because we can always find a $\delta(\epsilon) > 0$ for the
choice of $\kappa^2 = \frac{P^2}{(P+\sigma^2)\sigma_z^2} -
\frac{\epsilon}{\sigma_z^2}$. Hence from knowing $\mathbf{v'}$,
$\uv$ and $\yv$
 we can make an estimate of $\vv$.
 Since all random variables are
 Gaussian, the optimal estimate is a linear MMSE estimate which can
 be computed as follows.

Let $\mathbf{\Lambda}$ be the covariance matrix of $[V' \ U \
Y]^T$ and let $\mathbf{\Gamma}$ be the correlation between $V$ and
$[V' \ U \ Y]^T$. $\mathbf{\Lambda}$ and $\mathbf{\Gamma}$ are
given by

\[
 \mathbf{\Lambda} = \left(
                       \begin{array}{ccc}
                         \sigma_v^2-\sigma_z^2 & \kappa (\sigma_v^2-\sigma_z^2) & 0 \\
                         \kappa (\sigma_v^2-\sigma_z^2) & P +  \kappa^2 \sigma_v^2 & P  \\
                         0 & P  & P  + \sigma^2 \\
                       \end{array}
                     \right) \ \mbox{and} \ \mathbf{\Gamma} = \left(
                                                   \begin{array}{ccc}
                                                     \sigma_v^2-\sigma_z^2 & \kappa \sigma_v^2 & 0 \\
                                                   \end{array}
                                                 \right)^T
                                                 \].

The coefficients of the linear MMSE estimate are given by
$\mathbf{{\Lambda^{-1}}}\mathbf{\Gamma}$ and this yields the
optimal MMSE estimate which is given below as,

\begin{equation}\label{eqn:estimationwzhda} \hat{\vv} = \mathbf{v'} +
\frac{\kappa \sigma_z^2}{P}(\uv - \kappa \mathbf{v'} -\alpha \yv)
\end{equation}

The distortion $D$ is given by

\begin{eqnarray}
\label{eqn:disthdaww}
D &=& E[(\mathbf{v}-\mathbf{\hat{v}})^2] \nonumber \\
  &=& E[(\mathbf{v} - \mathbf{v'} -
\frac{\kappa \sigma_z^2}{P}(\uv - \kappa \mathbf{v'} -\alpha
\yv))^2] \nonumber \\
&=& E[(\mathbf{z}  - \frac{\kappa \sigma_z^2}{P}(\kappa \mathbf{z}
+\xv -\alpha \yv) )^2] \nonumber \\
&=& E[((1-\frac{\kappa^2 \sigma_z^2}{P}) \zv - \frac{\kappa
\sigma_z^2}{P}((1-\alpha)\xv -\alpha \wv))^2 ] \nonumber \\
&\stackrel{(a)}{=}& \frac{\sigma_z^2}{1+ \frac{P}{\sigma^2}} + \delta(\epsilon)\nonumber \\
\end{eqnarray}

Here, (a) follows by using the appropriate values of $\kappa$ and
$\alpha$. We once again obtain the optimal distortion $D$ by making
$\epsilon$ arbitrarily small and $\delta(\epsilon) \rightarrow 0$.
\end{enumerate}

It is instructive to compare the performance of this scheme with
that of the following naive scheme that would be optimal in the
absence of side-information at the receiver. In the naive scheme,
the $\vv$ is transmitted directly (analog transmission).  At the
receiver, an MMSE estimate of $\vv$ is formed from the received
signal $\yv$ and the available side information $\vv'$. The
distortion for this naive scheme can be seen to be $D_{naive} =
\sigma_z^2/(1 + (P/\sigma^2) \sigma^2_z)$.

Notice that $\frac{\partial D_{naive}}{\partial
\sigma_z^2}|_{\sigma_z^2 = 0} = 1$, whereas for the Wyner-Ziv
scheme, $\frac{\partial D}{\partial \sigma_z^2}|_{\sigma_z^2 = 0}
= \frac{1}{1 + P/\sigma^2} < 1$. At $\sigma_z^2 = 0$, both
$D_{naive}$ and $D$ are zero. i.e. the optimal scheme and the
naive scheme approach zero distortion with different slopes.
% making
%the optimal scheme arbitrarily better than the naive scheme in
%approaching zero distortion.

\subsection{Superimposed digital and HDA Wyner-Ziv scheme}

\begin{figure}[htbp]
\begin{center}
\includegraphics[scale=0.45,angle=0]{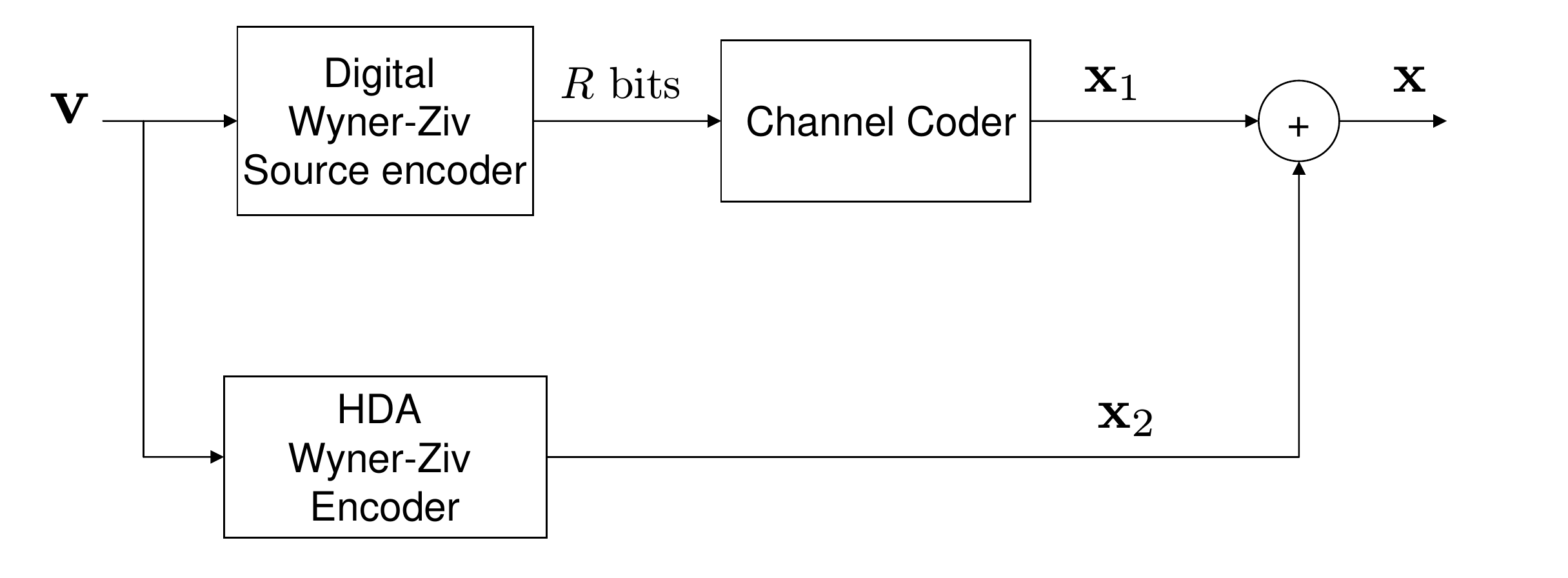}
\end{center}
\caption{Block diagram of the encoder of the superimposed digital
and HDA Wyner-Ziv scheme.} \label{fig:superimpdigiwz}
\end{figure}

 The above results could also be extended to a
form of superimposed digital and analog coding. This is similar to
the superimposed digital and HDA Costa coding case discussed in
section \ref{subsec:superimp digi and hda costa}. We once again have
two streams as shown in Fig~\ref{fig:superimpdigiwz}. The first
stream uses a rate $R$ Wyner Ziv code to quantize the source
assuming the side information $\mathbf{v'}$ is known at the receiver
and the discrete index is encoded using an optimal channel code to
produce the codeword $\xv_1$. The power allocated to this stream is
$P_{WZ} = (P+\sigma^2)(1-2^{-2R})$. The second stream uses the HDA
Wyner-Ziv scheme and produces the output $\xv_2$. The auxiliary
random variable of the HDA scheme is given by \begin{equation}
\label{eqn:auxiliaryrvhc} U = \kappa_1 V +  X_2
\end{equation}
 with $X_2 \sim {\mathcal {N}}(0,P_{HWZ})$, where $P_{HWZ} =
(P+\sigma^2)2^{-2R}-\sigma^2$  and $X_2$ and $V$ are independent. We
also choose $\kappa_1^2 =
\frac{P_{HWZ}^2}{(P_{HWZ}+\sigma^2)\sigma_e^2}-
\frac{\epsilon}{\sigma_e^2}$ where $ \sigma_e^2 =
\sigma_z^22^{-2R}$.

 The two streams ($\xv_1$ and $\xv_2$)
are superimposed and transmitted through the channel. The received
signal is given by $\yv = \xv_1 + \xv_2 + \wv$. At the receiver
$\xv_1$ is decoded assuming $\xv_2 + \wv$ as independent noise and
this gives the Wyner-Ziv encoded bits (index). This along with the
side information $\mathbf{v'}$ can be used to make an estimate of
the source $\vv$ and we call the estimate as $\mathbf{\tilde{v}}$.
The random variables corresponding to $\vv$ and $\mathbf{\tilde{v}}$
are related as
\begin{equation} \label{eqn:wzequivchannel}V = \tilde{V}+
\tilde{Z}\end{equation} with $\tilde{Z}$ having a variance
$\sigma_z^22^{-2R}$. When the digital part is first decoded and
canceled from the received signal, we get an equivalent channel for
the HDA Wyner-Ziv scheme with power constraint $P_{HWZ}$ and channel
noise $\sigma^2$. We next make a final estimate of $\vv$ using a HDA
Wyner-Ziv decoder from the new side information $\tilde{\vv}$, the
observed equivalent channel $(\yv-\xv_1)$ and the decoded $\uv$.
Notice that since the choice of
$\kappa_1^2 =
\frac{P_{HWZ}^2}{(P_{HWZ}+\sigma^2)\sigma_e^2}-
\frac{\epsilon}{\sigma_e^2}$ where $ \sigma_e^2 =
\sigma_z^22^{-2R}$ is
designed for the side information $\tilde{\vv}$, this ensures
decoding of $\uv$ with arbitrarily high probability.
 The achievable distortion is then given as follows.

\begin{eqnarray}
\label{eqn:distforsupwz}
D & \stackrel{(a)}{=} & \sigma_e^2\frac{\sigma^2}{P_{HWZ} + \sigma^2} + \delta(\epsilon) \nonumber\\
    &=& \sigma_z^22^{-2R}\frac{\sigma^2}{P_{HWZ} + \sigma^2} + \delta(\epsilon) \nonumber\\
         &\stackrel{(b)}{=}& \sigma_z^2 \frac{P_{HWZ} +\sigma^2}{P+\sigma^2} \frac{\sigma^2}{P_{HWZ} + \sigma^2} + \delta(\epsilon) \nonumber \\
         &=& \frac{\sigma_z^2}{1+\frac{P}{\sigma^2}} + \delta(\epsilon)\nonumber\\
\end{eqnarray}

Here in (\ref{eqn:distforsupwz}) (a) follows since we assume that
the first stream is decoded with high probability and apply the
results of HDA Wyner-Ziv decoding with the side information
$\mathbf{\tilde{v}}$. Also (b) follows since $P_{HWZ} =
(P+\sigma^2)2^{-2R}-\sigma^2$. The optimal distortion
$\frac{\sigma_z^2}{1+\frac{P}{\sigma^2}} $ can be obtained by making
$\epsilon$ arbitrarily small and $\delta(\epsilon) \rightarrow 0$.
Notice that for any rate $R$, $0 \leq R < C$, where $C$ is the
capacity of the AWGN channel, there is a corresponding power
allocation for $P_{HWZ} = (P+\sigma^2)2^{-2R}-\sigma^2$ for which
the overall scheme is optimal. Thus, there are infinitely many
schemes which are optimal with the digital Wyner-Ziv corresponding
to $P_{HWZ}=0$ and the HDA Wyner-Ziv corresponding to $P_{HWZ}=P$
and $R=0$.

Further, we would like to mention that there is another way to get a
family of optimal schemes using the HDA Wyner-Ziv scheme. Here, the
source $\vv$ is encoded using a HDA Wyner-Ziv encoder to the
sequence $\xv$. The auxiliary random variable $U$ is given by
\begin{equation}
U = \kappa V + X
\end{equation}
where $\kappa^2 = \frac{P^2}{(P+\sigma^2) \sigma_v^2} -
\frac{\epsilon}{\sigma_v^2}$.
 The sequence $\xv$ can be treated as an i.i.d
Gaussian source and, hence, the family of schemes proposed by Bross,
Lapidoth and Tinguely \cite{shraga06} can be applied on $\xv$.  The
scheme proposed in \cite{shraga06} quantizes the analog source,
which in this case is $\xv$ to a quantization index and is sent over
the Gaussian channel along with the uncoded analog source (here
$\xv$) with the appropriate power scaling. At the receiver we can
obtain an optimal estimate of $\xv$ by first decoding the quantized
index and then making an estimate on the analog source.  Notice that
the HDA Wyner-Ziv receiver only requires an optimal MMSE estimate of
$\xv$, which can be obtained using the family of schemes in
\cite{shraga06}. Hence the resulting distortion in $\vv$ is still
optimal. To establish this claim we need to show that $\uv$ can be
decoded with arbitrarily high probability and an optimal estimate of
$\vv$ must be made using $\uv$ and the MMSE estimate
$\mathbf{\hat{x}}$.

 We next show below that $I(U;V)
< I(U;V',\hat{X})$. Hence, we can choose a codebook for $\uv$ with
$2^{nR_1}$ codewords such that $I(U;V) < R_1 < I(U;V',\hat{X})$. Since
$I(U;V) < R_1$, we can find a $\uv$ that is jointly typical with $\vv$ with
probability close to 1 and since $R_1 < I(U;V',\hat{X})$, $\uv$ can decoded
with high probability from $(V',\hat{\xv})$.

\begin{eqnarray} \label{eqn:decodegenwz}
I(U;V',\hat{X}) & = & h(U) - h(U|V',\hat{X}) \nonumber\\
    &=& h(U) - h(U-\kappa V' - \hat{X}|V',\hat{X}) \nonumber \\
         &=& h(U) - h(\kappa Z +  X - \hat{X}  |\hat{X},V') \nonumber \\
         &\stackrel{(a)}{=}& h(U) - h(\kappa Z + X - \hat{X}) \nonumber \\
         &\stackrel{(b)}{=}& \frac{1}{2} \log \left(\frac{P + \kappa^2 \sigma_v^2}{\kappa^2
         \sigma_z^2 + \alpha\sigma^2}\right) \nonumber \\
         &{=}& \frac{1}{2} \log \left(\frac{P + \kappa^2\sigma_v^2}{P}\right) + \delta(\epsilon) \nonumber\\
         &=& h(U) - h(U|V) + \delta(\epsilon) \nonumber\\
         &=& I(U;V) + \delta(\epsilon)
\end{eqnarray}

In (\ref{eqn:decodegenwz}), (a) follows because  $(X - \hat{X})$ is
orthogonal to $\hat{X}$ and hence  $(\kappa Z + X - \hat{X})$ is
independent of $\hat{X}$ and $V'$, (b) follows because $X-\hat{X}$ is Gaussian with variance $\alpha \sigma^2$
and is orthogonal to $Z$.
 The estimate of $\vv$ is then given by
 \begin{equation}\hat{\vv} = \mathbf{v'}+\frac{\kappa \sigma_z^2}{P}(\uv-\kappa \vv' - \mathbf{\hat{x}})
  \end{equation}
The resulting distortion can be obtained by following the steps
similar to those in (\ref{eqn:disthdaww}) which can be found to be
optimal.

\section{Transmission of a Gaussian Source with Interference at the
Transmitter and Side Information at the
Receiver}\label{sec:combined}

In this section, we consider the problem of transmitting a Gaussian
source $\vv$ through an AWGN channel with channel noise variance
$\sigma^2$ in the presence of an interference $\sv$ known only at
the transmitter and in the presence of side information
$\mathbf{v'}$ known only at the receiver. The side information
$\mathbf{v'}$ is assumed to be related to the source $\vv$ according
to
\[
V = V' + Z
\]
where $Z \sim {\cal N}(0,\sigma_z^2)$ and is independent of $V'$.

A similar model has been considered by Merhav and Shamai
\cite{merhav03} for a more general setup where the source and side
information are not assumed to be Gaussian. They show that a
separation based approach of Wyner-Ziv coding followed by
Gelfand-Pinsker coding is optimal. Here, we propose a joint-source
channel coding scheme when the source and channel noise are
Gaussian. The proposed scheme is easily obtained by combining the
results from the previous two sections. It must be noted that a
similar joint source channel coding scheme using nested lattices
and dither has been shown in \cite{kochman06}. However, our scheme
is based only on random code books.

To establish our scheme we can combine the results from the previous
two sections as follows. Choose the auxiliary random variable $U$
such that
\begin{equation}
 U = X + \alpha S +\kappa V
 \end{equation} with $\kappa^2 =
\frac{P^2}{(P+\sigma^2)\sigma_z^2}-\frac{\epsilon}{\sigma_z^2}$ and
$\alpha = \frac{P}{P+\sigma^2}$. Further, let $X \sim
\mathcal{N}(0,P)$, $S\sim \mathcal{N}(0,Q)$ and $V\sim
\mathcal{N}(0,\sigma_v^2)$ and let $X$, $S$ and $V$ be  pairwise
independent. A codebook $\cal{U}$ is obtained by generating
$2^{nR_1}$ code sequences for $\uv$ and this is shared between the
encoder and decoder. At the encoder, the source $\vv$ is encoded by
choosing an $\xv$ that is jointly typical with $\uv$,$\vv$ and
$\sv$. Such a $\uv$ exists with high probability if we have chosen
$R_1 > I(U;S,V) $. Now $\xv$ is transmitted over the channel. The
received signal vector $\yv$ is given as
\[
\yv = \xv + \sv + \wv
\]

At the decoder, $\uv$ is decoded by looking for a $\uv$ that is
jointly typical with $\yv$ and the side information $\mathbf{v'}$.
Using standard arguments on joint-typicality, it can be seen that
a unique such $\uv$ exists with high probability if $R_1 <
I(U;Y,V')$. We now show that $I(U;S,V) < I(U;Y,V')$. This implies
that there exists an $R_1$, such that $I(U;S,V) < R_1 < I(U;Y,V')$
which satisfies the requirements at the encoder and the decoder.

\begin{eqnarray}
\label{eqn:wzandcosta}
I(U;Y,V') &=& h(U) - h(U|Y,V') \nonumber \\
          &=& h(U) - h(U-\alpha Y -\kappa V'|Y,V') \nonumber \\
          &=& h(U) - h(\kappa Z + (1-\alpha)X - \alpha W|Y,V') \nonumber\\
          &\stackrel{(a)}{=}& h(U) - h(\kappa Z + (1-\alpha)X - \alpha W) \nonumber          \\
          &=& \frac{1}{2} \log(\frac{P + \alpha^2 Q + \kappa^2
          \sigma_v^2}{P}) + \delta(\epsilon) \nonumber \\
          &=& I(U;S,V) + \delta(\epsilon)
\end{eqnarray}

where (a) follows since $\kappa Z + (1-\alpha)X - \alpha W$ is
orthogonal to $Y$ and $V'$. Then an optimal linear MMSE estimate
of $\vv$ is formed from the side information $\mathbf{v'}$, the
received vector $\yv$ and the vector $\uv$. By using the argument
as in section.~\ref{subsec:hdawz}, the MMSE estimate is given by
\begin{equation}
\hat{\vv} = \mathbf{v'}+\frac{\kappa \sigma_z^2}{P}(\uv-\kappa \vv'
-\alpha \yv)
  \end{equation}

  The resulting distortion can be obtained by following steps
  similar to (\ref{eqn:disthdaww}) and can be seen to be $D = \frac{\sigma_z^2}{1+\frac{P}{\sigma^2}}$,
  which is the optimal distortion.

\section{Analysis of the schemes for SNR mismatch}\label{sec:analysisSNRmismatch}

In this section, we consider the performance of the above JSCC
schemes for the case of SNR mismatch where we design the scheme to
be optimal for a channel noise variance of $\sigma^2$, but the
actual noise variance is $\sigma_a^2$.

Separation based digital schemes suffer from a pronounced
threshold effect. When the channel SNR is worse than the designed
SNR, the index cannot be decoded and when the channel SNR is
better than the designed SNR, the distortion is limited by the
quantization and does not improve. However, the hybrid digital
analog schemes considered offer better performance in this
situation.

Let us consider the joint source channel coding setup with side
information at both the transmitter and receiver and $\sigma_a^2 < \sigma^2$. We can decode
$\uv$ at the receiver when the SNR is better than the designed SNR
and make an estimate of the source from the various observations
at the receiver as shown below.

\begin{figure}[htbp]
\begin{center}
\includegraphics[scale=0.65,angle = 0]{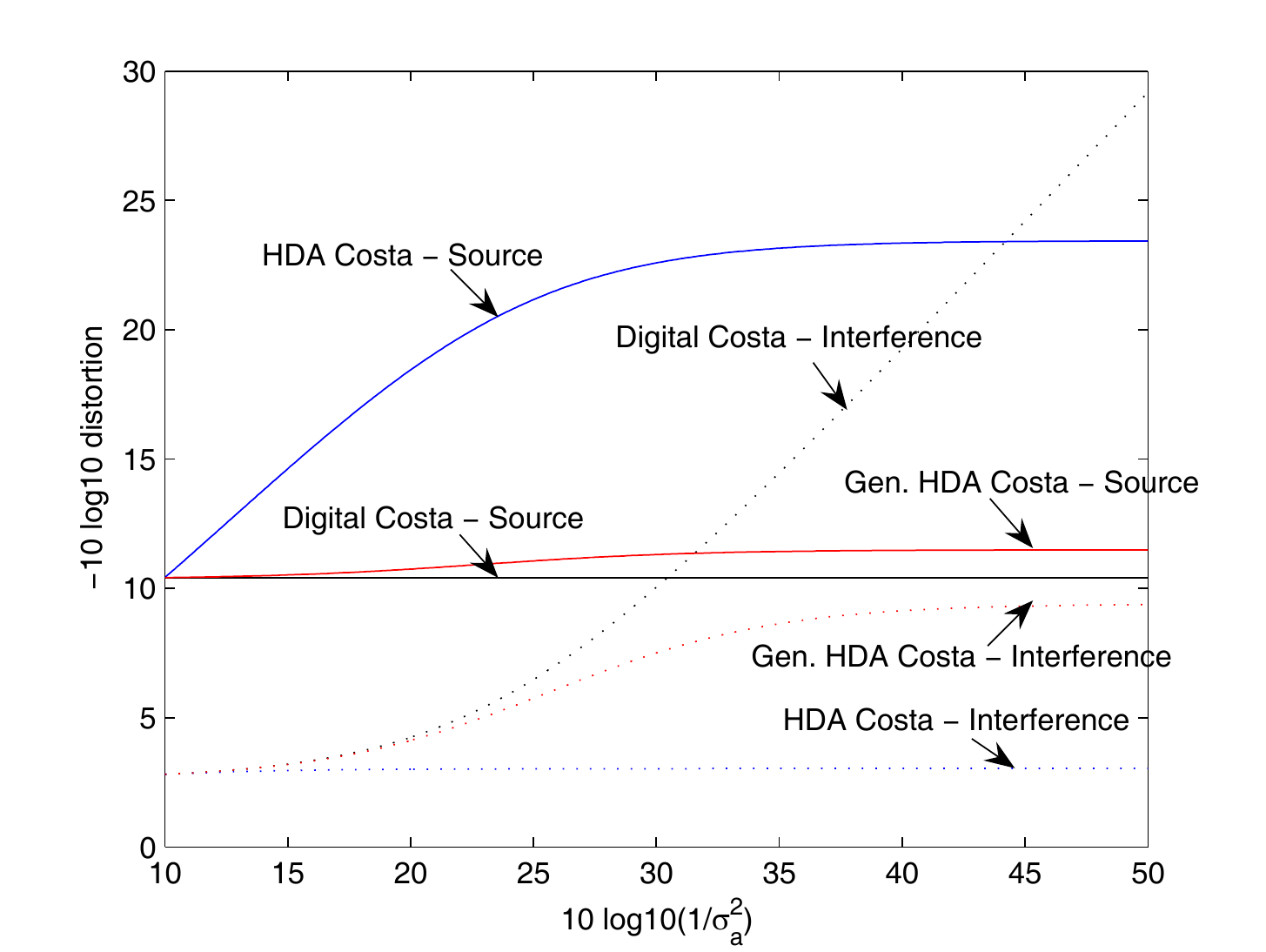}
\end{center}
\caption{Performance of the different Costa coding schemes for the
joint source channel coding problem.}
\label{fig:Diganahybperformance}
\end{figure}

\begin{equation}
U = X + \alpha S + \kappa_w V \label{eqn:auxrvU2sidedinfo}
\end{equation}
\begin{equation}
V = V' + Z \label{eqn:auxrvV2sidedinfo}
\end{equation}
\begin{equation}
Y = X + S + W_a \label{eqn:auxrvY2sidedinfo}
\end{equation}

where $\kappa_w = \sqrt{\frac{P^2}{(P+\sigma^2)\sigma_z^2}}$,
$\alpha = \frac{P}{P+\sigma^2}$, $S \sim \mathcal{N}(0,Q)$ and $Z
\sim \mathcal{N}(0,\sigma_z^2)$. From now on, we drop the $\epsilon$'s
in $\kappa_w$ to improve clarity. Note that $\alpha$ depends only
on the assumed noise variance $\sigma^2$ and not on $\sigma_a^2$.

From the observations $[V',U,Y]$, an optimal linear MMSE estimate
of $V$ is obtained. Similar to the definition in
section~\ref{subsec:hdawz} let $\mathbf{\Lambda}$ be the
covariance of $[V',U,Y]^T$ and ${\mathbf{\Gamma}}$ be the
correlation between $V$ and $[V',U,Y]^T$.

Hence\[ \mathbf{\Lambda} = \left(
                       \begin{array}{ccc}
                         \sigma_v^2-\sigma_z^2 & \kappa (\sigma_v^2-\sigma_z^2) & 0 \\
                         \kappa (\sigma_v^2-\sigma_z^2) & P + \alpha^2 Q + \kappa_w^2 \sigma_v^2 & P + \alpha Q \\
                         0 & P + \alpha Q & P + Q + \sigma_a^2 \\
                       \end{array}
                     \right) \ \mbox{and} \ \mathbf{\Gamma} = \left(
                                                   \begin{array}{ccc}
                                                     \sigma_v^2-\sigma_z^2 & \kappa \sigma_v^2 & 0 \\
                                                   \end{array}
                                                 \right)^T.\]

Then the distortion (in the presence of mismatch) is given by
\begin{equation}
D_a = \sigma_v^2 - \mathbf{\Gamma}^T
\mathbf{\Lambda}^{-1}\mathbf{\Gamma}
\end{equation}

This on further simplification yields

\begin{equation}
\begin{split}
& D_a = {\left[(Q \sigma^4 + (P(P+Q)+2P\sigma^2 +
\sigma^4)\sigma_a^2)\sigma_z^2\right]}\times \\ &\left[P^2(P+Q) +
P(P+Q)\sigma^2 + Q \sigma^4 + (P(2P+Q)+
3P\sigma^2+\sigma^4)\sigma_a^2\right]^{-1}.
\end{split}
\label{eqn:comprehenall}
\end{equation}

Let us now look at a few special cases

\subsection {Hybrid Digital Analog Costa Coding} In this setup there is side information
only at the transmitter. The distortion achievable for the user
under SNR mismatch with the actual SNR greater than the designed SNR
 is obtained by setting $\sigma_v = \sigma_z$
(\ref{eqn:comprehenall}) and is given below.

\begin{equation}
\label{eqn:HDACostaDistexp}
\begin{split}
& D_{va} = {\left[(Q \sigma^4 + (P(P+Q)+2P\sigma^2 +
\sigma^4)\sigma_a^2)\sigma_v^2\right]}\times \\ & \left[P^2(P+Q) +
P(P+Q)\sigma^2 + Q \sigma^4 + (P(2P+Q)+
3P\sigma^2+\sigma^4)\sigma_a^2\right]^{-1}
\end{split}
\end{equation}

The distortion in the source $\vv$ is shown in
Fig.\ref{fig:Diganahybperformance} for a designed SNR of 10~dB as
the actual channel SNR ($10 \log 1/\sigma_a^2$) varies when the
source and interference both have unit variance. It can be seen that
the distortion in the source is smaller with the HDA Costa scheme
than with the digital Costa scheme.

%It is interesting to see the performance when the $SNR \rightarrow
%\infty$ given by,
%
%\begin{equation}
%D_{va} = \frac{Q \sigma^4 \sigma_v^2}{P^3 + PQ\sigma^2 + Q \sigma^4
%+ P^2(Q+\sigma^2)}
%\end{equation}
%This shows that this scheme has a distortion exponent  of $0$.

In some case, the distortion in estimating the interference at the
receiver may also be of interest and can be obtained by estimating
$S$ from (\ref{eqn:auxrvU2sidedinfo}) and
(\ref{eqn:auxrvY2sidedinfo}). The distortion is given below,
\begin{equation}
\begin{split}
 & D_{sa} = {\left[Q(P+\sigma^2)(P^2+(2P+\sigma^2)\sigma_a^2)\right]}
\times \\ & \left[P^2(P+Q) + P(P+Q)\sigma^2 + Q \sigma^4  (P(2P+Q)+
3P\sigma^2+\sigma^4)\sigma_a^2\right]^{-1}
\end{split}
\end{equation}

%We again analyze the performance when the $SNR \rightarrow \infty$,
%which yields,
%
%\begin{equation}
%D_{sa} = \frac{Q(P+\sigma^2)P^2}{P^2(P+Q) + P(P+Q)\sigma^2 + Q
%\sigma^4 }
%\end{equation}
%The distortion exponent again is $0$.

It can be seen from Fig.~\ref{fig:Diganahybperformance} that the
distortion in estimating the interference is better for the digital
scheme than for the HDA Costa scheme.

In \cite{kimcover}, Sutivong {\em et al.} have studied a somewhat
related problem. They consider the transmission of a digital
source in the presence of an interference known at the transmitter
with a fixed channel SNR. They study the optimal tradeoff between
the achievable rate and the error in estimating the interference
at the designed SNR. The main result is that we can get a better
estimate of the interference if we transmit the digital source at
a rate lesser than the channel capacity. There are important
differences our work and that in \cite{kimcover}. First of all, we
consider transmission of an analog source instead of a digital
source. Secondly, we consider mismatch in the channel, i.e., our
schemes are designed to be optimal at the designed SNR and as we
move away from the designed SNR, we study the tradeoff between the
error in estimating the interference and the distortion in the
reconstruction of the analog source. This tradeoff is discussed
below.

%This can be seen in Fig.~\ref{fig:Diganahybperformance}. The
%interference and the source have unit variance and the different
%schemes are designed to be optimal at an SNR of 10db and we study
%the performance of the schemes when the actual SNR turns to be
%different.

\subsection{Generalized HDA Costa Coding under channel mismatch}

 Next we analyze the performance of the generalized HDA Costa coding under channel
 mismatch. This case leads to some interesting analysis.
 By changing the source coding rate of the digital part $R$, we
 can tradeoff the distortion between the source and the
 interference in the presence of mismatch.

The
 different random variables and their relations are given below.

%\begin{figure}[htbp]
%\begin{center}
%\epsfxsize=3.5 in \epsfbox{costatradeoff.eps}
%\end{center}
%\caption{Tradeoff between interference and source distortion at 50
%db. The hybrid schemes are designed to be optimal at 40db.}
%\label{fig:costatradeoff}
%\end{figure}

 \begin{equation}
U = X +\alpha S + \kappa_1 E
 \end{equation}

\begin{equation}
Y = X + S + W_a
\end{equation}

\begin{equation}
V = V^* + E
\end{equation}

In the above equation $\kappa_1 =
\sqrt{\frac{P}{P+\sigma^2}\frac{(P+\sigma^2)-\sigma^22^{2R}}{\sigma_v^2}}$
(Again, we have dropped the $\epsilon$ in the expression for $\kappa_1$.)
From the above equations an estimate of $S$  as well as $V$ is
obtained by taking a linear MMSE estimate as all the random
variables are Gaussian. The resulting expressions of estimation
error $D_{sa}(R)$ and $D_{va}(R)$ are given by

\begin{equation}
\begin{split}
 & D_{va}(R) =
{\left[(\sigma_a^2(\sigma^2+P)^2+(\sigma^4+\sigma_a^2P)Q)\sigma_v^2\right]}
\times
\\ & {\left[(\sigma^2 + P)^2(\sigma_a^2 + P +
Q)-2^{2R}(\sigma^2-\sigma_a^2)P(\sigma^2+P+Q)\right]^{-1}}
\end{split}
\end{equation}

\begin{equation}
\begin{split}
& D_{sa}(R) =
{\left[(\sigma^2+P)(2^{2R}(\sigma^2-\sigma_a^2)P-(\sigma^2 +
P)(\sigma_a^2 + P )) Q\right]} \\ & \times
{\left[2^{2R}(\sigma^2-\sigma_a^2)P(\sigma^2+P+Q)-(\sigma^2 +
P)^2(\sigma_a^2 + P + Q)\right]^{-1}}
\end{split}
\end{equation}

The performance of the generalized HDA Costa scheme and HDA Costa
scheme in relation to digital scheme is shown in
fig.~\ref{fig:Diganahybperformance}.  For example in separation
using digital Costa there is no improvement in our estimate of the
analog source, but we get a better estimate of the interference as
shown in fig.\ref{fig:Diganahybperformance}. On the contrary for the
HDA Costa scheme there is only a small improvement in the estimate
of the interference but a good improvement in the estimate of the
analog source.  The generalized HDA also shows a difference in the
estimate for the source and the interference for different rates $R$
and performs as a digital Costa for the choice of $R=C$ and as HDA
Costa for the choice of $R=0$. In effect we can tradeoff the
estimation error in interference with the source by choosing
different values of $R$ when there is a channel mismatch.

%\underline{what is the relationship to Sutivong's paper?}
%
%\underline{do we need the rest of this section?}

%However to understand the tradeoff between the two schemes we
%calculate the slope of $D_{sa}(R)$ with respect $D_{va}(R)$.
%\begin{equation}
%\frac{dD_{sa}}{dD_{va}} =
%-\frac{(\sigma^2-\sigma_a^2)(\sigma^2+P)^2Q^2}{(\sigma^2+P+Q)(\sigma_a^2(\sigma^2+P)^2+(\sigma^4+\sigma_a^2P)Q)\sigma_v^2}
%\end{equation}
%
%The slope hence turns out to be a constant. This shows that the same
%results could be obtained by using time sharing between Analog and
%Digital Costa even though the hybrid Costa scheme appears different.
%This is illustrated in Fig.~\ref{fig:costatradeoff}.

%\includegraphics[height=120mm]{broadcastvscosta5.eps}

\subsection{Hybrid Digital Analog Wyner Ziv}
In this case the distortion could be obtained by setting $Q=0$ in
(\ref{eqn:comprehenall}). The actual distortion is given by

\begin{figure}[htbp]
\begin{center}
\includegraphics[scale=0.65,angle = 0]{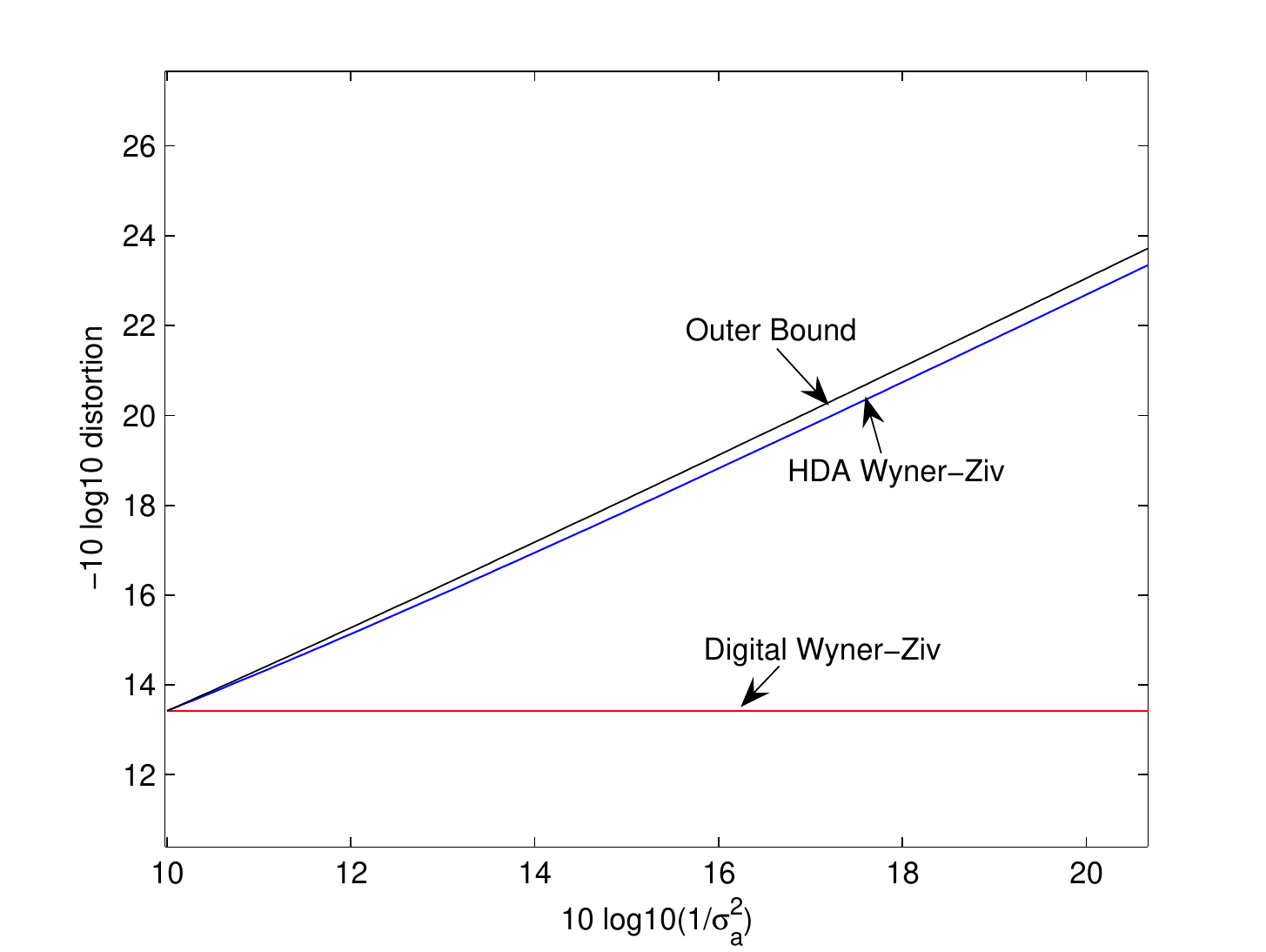}
\end{center}
\caption{Performance of the different Wyner-Ziv schemes for the
joint source channel coding problem.}
\label{fig:Diganawynerzivperformance}
\end{figure}

\begin{equation}\label{eqn:HDAWZdistexpression}
D_{a} = \frac{(P+\sigma^2)\sigma_a^2 \sigma_z^2}{P^2 + (2P
+\sigma^2)\sigma_a^2}
\end{equation}

%In the Wyner-Ziv case we see the distortion exponent is -1 as the
%distortion decays as an order of $SNR^{-1}$. A better analysis gives
%a more interesting observation.

This is clearly better than $\frac{\sigma_z^2 \sigma^2}{P +
\sigma^2}$ which is what is achievable with a separation based
approach. However, we don't know if this is the optimal distortion
that is achievable in the presence of channel mismatch. A simple
lower bound on the achievable distortion in the presence of
mismatch is to assume that the transmitter knows the channel SNR.
Based on this we can analyze the gap in dB between the distorion
of HDA Wyner Ziv scheme and the lower bound as follows.

The lower bound on $D$ is given by
\begin{equation}
D_{lower} = \frac{\sigma_z^2}{1 + P/\sigma_a^2}
\end{equation}

Now the gap between the analog Wyner-Ziv and the bound at high SNR
can be easily calculated as $\lim_{\sigma_a \rightarrow 0}
\frac{D_{lower}}{D_a}$. The gap in db, $G$ is hence given by

\begin{equation}
G = 10 \log \left(\frac{P}{P + \sigma^2}\right)
\end{equation}
 For example, if our designed
SNR is say 10 db, for high SNRs, we loose at most $G = 0.41$ db
which is fairly close to the outer bound as shown in
Fig.~\ref{fig:Diganawynerzivperformance}.

\section{Distortion exponent for HDA Costa and Wyner-Ziv schemes}
\label{sec:distexp}

In this section, we consider the performance of the HDA joint
source-channel coding schemes for transmitting a Gaussian source
through a Gaussian channel when the actual channel noise variance
$\sigma_a^2$ is not known, but it is known that the variance is
always smaller than $\sigma^2$. Since we are interested in the
performance of a single encoding scheme over a wide range of noise
variances, a useful measure of performance is the rate of decay of
the distortion as a function of the actual noise variance in the
limit $\sigma_a^2 \rightarrow 0$. More precisely, we define a
distortion exponent as
\[
\zeta = \lim_{\sigma_a^2 \rightarrow 0} \frac{\log (D({
\sigma_a^2}))}{\log({\sigma_a^2})}
\]
where $D({\sigma_a^2})$ is the distortion when the noise variance
is $\sigma_a^2$. Notice that this exponent is quite different from
the distortion signal-to-noise ratio (SNR) exponent considered in
\cite{laneman05}, \cite{holliday04} for the case of slow fading
channels. In \cite{laneman05} and \cite{holliday04}, the rate of
decay of distortion with average SNR is studied by allowing for a
family of coding schemes, one for each average SNR. In contrast,
we fix the encoding scheme here and consider the rate of decay
with the actual channel SNR (i.e., there is no fading).

An upper bound on the achievable $\zeta$ can be obtained by
assuming that a genie informs the transmitter of ${\sigma_a}$ and
the transmitter chooses an optimal encoding scheme for a noise
variance of $\sigma_a^2$. Let us assume a general case where there
is an interference $\sv$ ($S \sim {\cal N}(0,Q)$) which is known
at the transmitter and some side information $\vv'$ which is
related to $\vv$ according to $V = V' + Z$, where $Z \sim
{\cal N}(0,\sigma_z^2)$ is known at the receiver. Then, the
distortion for the genie-aided scheme is
$\frac{\sigma_z^2}{1+\frac{P}{\sigma_a^2}}$ since an optimal
Wyner-Ziv encoder followed by an optimal Costa encoder can be
chosen. In this case, the distortion exponent is 1.  Notice that
in the absence of any side information the distortion for the
genie-aided scheme is $\frac{\sigma_v^2}{1+\frac{P}{\sigma_a^2}}$
which also results in an exponent of 1. In the absence of any
interference also, the achievable distortion is
$\frac{\sigma_z^2}{1+\frac{P}{\sigma_a^2}}$ and the exponent is 1.
Thus, for any single encoding scheme, $\zeta \leq 1$ both in the
presence and absence of interference and/or side information.

We will now consider the performance of the HDA schemes considered
in Section~\ref{sec:combined}. If a joint source channel coding
scheme is designed to be optimal when the noise variance is
$\sigma^2$, then the distortion when the noise variance is
$\sigma_a^2$ is given by (from (\ref{eqn:comprehenall}) in
Section~\ref{sec:analysisSNRmismatch})
\begin{equation}
\begin{split}
& D({\sigma_a^2}) = {\left[(Q \sigma^4 + (P(P+Q)+2P\sigma^2 +
\sigma^4)\sigma_a^2)\sigma_z^2\right]}\times \\ &\left[P^2(P+Q) +
P(P+Q)\sigma^2 + Q \sigma^4 + (P(2P+Q)+
3P\sigma^2+\sigma^4)\sigma_a^2\right]^{-1}.
\end{split}
\label{eqn:comprehenall2}
\end{equation}
We now consider two cases.
\subsection{Absence of Interference}
When there is no interference at the transmitter, $Q=0$ and,
hence, from (\ref{eqn:comprehenall2}), we can see that the optimal
distortion can be obtained for a noise variance of $\sigma^2$ and
the optimal distortion exponent of $\zeta = 1$ can be obtained.
Thus, this scheme performs as well as the genie-aided receiver in
the distortion exponent sense.

\subsection{Presence of Interference}
In the presence of an interference, $Q \neq 0$, and from
(\ref{eqn:comprehenall2}), $\zeta$ can be seen to be zero. That
is, some amount of residual interference is always present and,
hence, in the high SNR limit $(\sigma_a^2 \rightarrow 0)$, the
performance is dominated by this residual interference. However,
if optimal performance is not desired when the noise variance is
$\sigma^2$, then the optimal exponent of $\zeta = 1$ can be
obtained using a minor modification to the scheme discussed in
Section~\ref{sec:combined}.  In the modified scheme, the auxiliary
random variable $U$ is generated as follows

\begin{equation}
U = X + S + \kappa_e V
\end{equation}
 Note that $\alpha$ is chosen
to be 1, which is clearly not optimal for a noise variance of
$\sigma^2$. The side information $V'$ is
\begin{equation}
V = V' + Z
\end{equation}

 and the received signal is
\begin{equation}
Y = X + S + W
\end{equation}

Using arguments similar to those in Section~\ref{sec:combined},
$\kappa_e$ is chosen so as to satisfy $I(U;Y,V')
> I(U;S,V )$. The required condition on $\kappa_e$ can be obtained
as follows
\begin{eqnarray}
\label{eqn:derivforke}
I(U;Y,V') &>& I(U;S,V) \nonumber \\
\Rightarrow h(U) - h(U|Y,V') &>& h(U) - h(U|S,V) \nonumber\\
\Rightarrow h(U|S,V) &>&  h(U|Y,V')
\end{eqnarray}

Note that $h(U|S,V) = h(X)$ and $h(U|Y,V') = h(U - \eta Y - \kappa_e
V' | Y,V')$, where $\eta = \frac{E[UY]}{E[Y^2]}$. For this choice of
$\eta$, $(U - \eta Y - \kappa_e V') \perp Y , V'$ and, hence,
$h(U|Y,V') = h(U - \eta Y - \kappa_e V')$. Hence, we get the
relation,
\begin{eqnarray}
\nonumber h(X) & > & h(U - \eta Y - \kappa_e V') \\
\nonumber \Rightarrow P  & > & E[(U-\eta Y - \kappa_e V')^2] \\
\nonumber \Rightarrow
\sqrt{\frac{P^2+PQ-Q\sigma^2}{(P+Q+\sigma^2)\sigma_z^2}} & > &
\kappa_e
\end{eqnarray}

Hence,
 $\kappa_e$ can be chosen to be arbitrarily close to
 $\sqrt{\frac{P^2+PQ-Q\sigma^2}{(P+Q+\sigma^2)\sigma_z^2}}$. Now $\mathbf{x}$ is transmitted
  and $\mathbf{y}$ is received. The optimal
 distortion is obtained as an MMSE estimate of $\mathbf{v}$ from
 $[\mathbf{y},\uv,\vv']$. The final distortion is given by

 \begin{equation}
D({\sigma_a^2}) = \frac{(P+Q)\sigma_a^2
\sigma_z^2}{(P+Q)\sigma_a^2 + \kappa_e^2(P+Q+\sigma_a^2)
\sigma_z^2}
 \end{equation}

It can be seen that as ${\sigma^2_a} \rightarrow 0$,
$D({\sigma_a^2}) \propto \sigma_a^2$ and, hence, $\zeta = 1$.

\section{Applications to Transmitting a Gaussian source with Bandwidth Compression}
\label{sec:gcbc}

We now consider the problem of transmitting $K$ samples of the
i.i.d Gaussian source to a single user in $N = \lambda K$ ($\lambda < 1$) uses of
an AWGN channel with noise variance $\sigma^2$ where the transmit power
is constrained to 1. There is no interference in the channel, but since $\lambda <
1$, we will see that the techniques described in the previous
sections are useful for this problem.

There at least three ways to achieve the optimal distortion in
this case. One is to use a conventional separation based approach.
The second one is to use superposition coding and the third one is
to use Costa coding. Although, they are all optimal for the single
user case, they perform differently when there is a mismatch in
the channel SNR and, hence, the last two approaches are briefly
described here.

\paragraph{Superposition Coding}

%\begin{figure}[htb]
%\begin{center}
%\center{\includegraphics[scale=0.5,angle = 0]{sysmodsup}}
%\caption{Encoder model using superposition coding for single user}
%\label{fig:supcod}
%\end{center}
%\end{figure}

Here we split the source in two parts and take $N$ samples of the
source $\vv$, namely $v_1^N$ and scale it by $\sqrt{a}$ creating the
systematic signal $\xv_1 = \sqrt{a} v_1^N$. We take the other $K-N$
source samples $v_{N+1}^K$ and use a conventional source encoder
followed by a capacity achieving channel code resulting in the $N$
dimensional vector $\xv_c = {\cal C} ( {\cal Q} (v_{N+1}^K))$, where
${\cal C}$ denotes a channel encoding operation and ${\cal Q}$
denotes a source encoding operation. Then $\xv_c$ is normalized so
that the average power is ${\sqrt{1-a}}$. The overall transmitted
signal is $\xv = \xv_s + \xv_c$ and the received signal is $\yv =
\xv + \wv$. At the receiver, the digital part is first decoded
assuming the systematic (analog) part is noise and then $\xv_c$ is
subtracted from $\yv$. Then an MMSE estimate of $v_1^N$ is formed.
For the optimal choice of $a$, the optimal overall distortion can be
obtained given by
\begin{equation}
\label{asupstar} a^{*}_{sup} = \sigma^2 \left[ \left(1+
\frac{1}{\sigma^2} \right)^\lambda - 1\right] {\mbox { and}} \
D^*_{sup} = \frac{1}{\left( 1+\frac {1} {\sigma^2}
\right)^\lambda}
\end{equation}
which is the optimal distortion.

\paragraph{Digital Costa Coding}
\begin{figure}[htb]
\begin{center}
\center{\includegraphics[scale=0.45,angle=0]{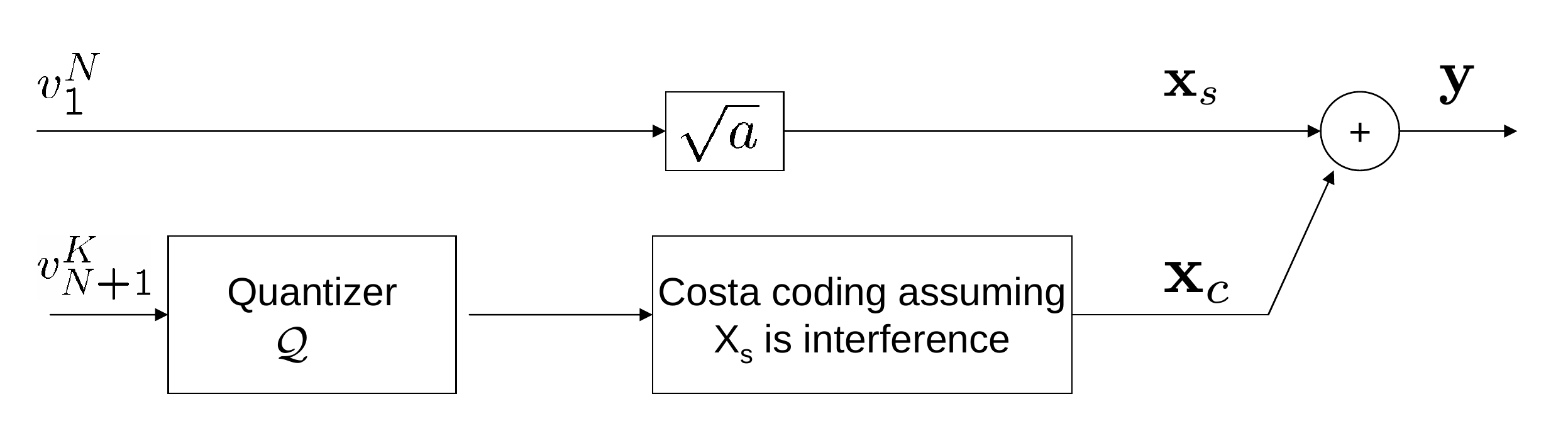}}
\caption{Encoder model using Costa coding for single user}
\label{fig:costacod}
\end{center}
\end{figure}
We split the source exactly as in the previous case and one stream
is formed as $\xv_s = \sqrt{a} v_1^N$. However, here the digital
part assumes that $\xv_s$ is interference and uses Costa coding
 to produce $\xv_c$ with power $1-a$ as shown in
Fig.~\ref{fig:costacod}. In Costa coding, we define an auxiliary
random variable $\uv = \xv_c + \alpha_1 \xv_s$ where $\alpha_1 =
\frac{1-a}{1-a+\sigma^2}$ is the optimum scaling coefficient. At
the receiver, the digital part is decoded which means that $\uv$
can be obtained. In spite of knowing $\uv$ exactly, the optimal
estimate of $v_1^N$ is obtained by simply treating $\xv_c$ as
noise since for the optimal choice of $\alpha_1$, $\xv_c = \uv -
\alpha_1 \xv_s$ and $v_1^N$ are uncorrelated. Therefore, an MMSE
estimate of $v_1^N$ is formed assuming $\xv_c$ were noise. Hence,
the overall distortion becomes

\begin{equation}
D = \frac{\lambda}{1+\frac{a}{1-a+\sigma^2}} +
\frac{1-\lambda}{\left( 1 +
\frac{1-a}{\sigma^2}\right)^{\lambda/1-\lambda}}
\end{equation}

Again, minimizing $D$ w.r.t. $a$ gives

\begin{equation}
\label{acostastar} a_{costa}^* = (1+\sigma^2) \left[ 1 -
\frac{1}{\left( 1 + \frac{1}{\sigma^2}\right)^\lambda}\right]
{\mbox {and}}  \ D^*_{costa} =\frac{1}{\left( 1+\frac {1}
{\sigma^2} \right)^\lambda} \end{equation} which is the best
possible distortion.

\paragraph{Hybrid Digital Analog Costa Coding} For the case of $\lambda = 0.5$, the digital
Costa coding part can be replaced by a hybrid digital analog (HDA)
Costa coding. We refer to such a scheme as HDA Costa coding. The
same power allocation however, remains the same and hence, we can
simply use $a^*_{Costa}$ without the need to differentiate the
digital and HDA Costa coding. It is quite straightforward to show
that $a^*_{Costa}
> a^*_{sup}$ for $\lambda < 1$. Hence, the Costa coding approach
allocates higher power to the systematic part than the
superposition approach, since the systematic part is treated as
interference.

\subsection{Performance in the presence of SNR mismatch}
\label{sec:bandcompmismatch}

Now, we consider the same set up as above, but when the actual
channel noise variance is $\sigma_a^2$, whereas the designed noise
variance is $\sigma^2$.

{\underline {Case 1:} $\sigma_a^2 > \sigma^2$}

The distortion for the superposition code can be computed to be
the sum of the distortions in the systematic part and the digital
part. When $\sigma_a^2 > \sigma^2$, the digital part cannot be
decoded and, hence, we assume that the distortion in the digital
part is the variance of the source, 1.

\begin{equation}
D_{sup} =
\frac{\lambda}{1+\frac{a^*_{sup}}{1-a^*_{sup}+\sigma_a^2}} +
(1-\lambda) \cdot 1
\end{equation}

Both the digital and HDA Costa coding schemes perform identically
when $\sigma_a^2 > \sigma^2$ and the distortion for the Costa
code can be computed to be
\begin{equation}
D_{digCosta} = D_{HDACosta} =
\frac{\lambda}{1+\frac{a^*_{Costa}}{1-a^*_{Costa}+\sigma_a^2}} +
(1-\lambda) \cdot 1
\end{equation}

 {\underline {Case 2:} $\sigma_a^2 < \sigma^2$}
In this case, the digital part can be decoded exactly and, hence,
the distortion for superposition coding is
\begin{equation}
D_{sup} = \lambda \ \frac{1}{1+\frac{a^*_{sup}}{\sigma_a^2}} +
(1-\lambda) \ \frac{1}{\left( 1 +
\frac{1-a^*_{sup}}{a^*_{sup}+\sigma^2}\right)^{\lambda/(1-\lambda)}}
\end{equation}
For digital Costa coding, the decoder first decodes the digital
part when the auxiliary random variable $\uv$ is perfectly known.
In the case when $\sigma_a^2 \neq \sigma^2$, the receiver must
form the MMSE estimate of $v_1^N$ from the channel observation
$\yv$ and $\uv$. Therefore, the overall distortion is
\begin{equation}
\begin{split}
& D_{digCosta} = \lambda \left(
1 - [\sqrt{a^*_{Costa}} \ \alpha \sqrt{a^*_{Costa}}]\times \right. \\ & \left. \left[%
\begin{array}{cc}
  1+\sigma_a^2 & 1-{a^*_{Costa}}+\alpha {a^*_{Costa}} \\
  1-a^*_{Costa}+\alpha a^*_{Costa} & 1-a^*_{Costa}+\alpha^2 a^*_{Costa} \\
\end{array}%
\right]^{-1} \times \right. \\ & \left.
\left[%
\begin{array}{c}
  \sqrt{a^*_{Costa}} \\
  \alpha \sqrt{a^*_{Costa}} \\
\end{array}%
\right] \right)
  +  (1-\lambda) \frac{1}{\left( 1 +
\frac{1-a^*_{Costa}}{a^*_{Costa}+\sigma^2}\right)^{\lambda/(1-\lambda)}}
\end{split}
\end{equation}

For the HDA Costa coding, we can decode $\uv$ and form MMSE
estimates of $v_{1}^N$ and $v_{N+1}^K$ separately and, hence, the
overall distortion is given by \hspace*{-0.5in}
\begin{equation}
\begin{split}
  & D_{HDACosta} =  \lambda \left(
1 - [\sqrt{a^*_{Costa}} \ \alpha \sqrt{a^*_{Costa}}] \times
\right.
\\ & \left.
 \left[
\begin{array}{cc}
  1+\sigma_a^2 & 1-{a^*_{Costa}}+\alpha {a^*_{Costa}} \\
  1-a^*_{Costa}+\alpha a^*_{Costa} & 1-a^*_{Costa}+\alpha^2 a^*_{Costa} + \kappa^2\\
\end{array}
\right]^{-1} \times \right. \\ & \left. \left[
\begin{array}{c}
  \sqrt{a^*_{Costa}} \\
  \alpha \sqrt{a^*_{Costa}} \\
\end{array}
\right] \right)
  +  (1-\lambda) \left (1 - [ 0 \ \kappa]\times \right. \\ & \left. \left[%
\begin{array}{cc}
  1+\sigma_a^2 & 1-{a^*_{Costa}}+\alpha {a^*_{Costa}} \\
  1-a^*_{Costa}+\alpha a^*_{Costa} & 1-a^*_{Costa}+\alpha^2 a^*_{Costa} + \kappa^2\\
\end{array}
\right]^{-1} \times \right. \\ & \left.  \left[
\begin{array}{c}
  0 \\
  \kappa \\
\end{array}
\right]
 \right)
 \end{split}
\end{equation}

\begin{figure}[htbp]
\begin{center}
\includegraphics[scale=0.5,angle = 90]{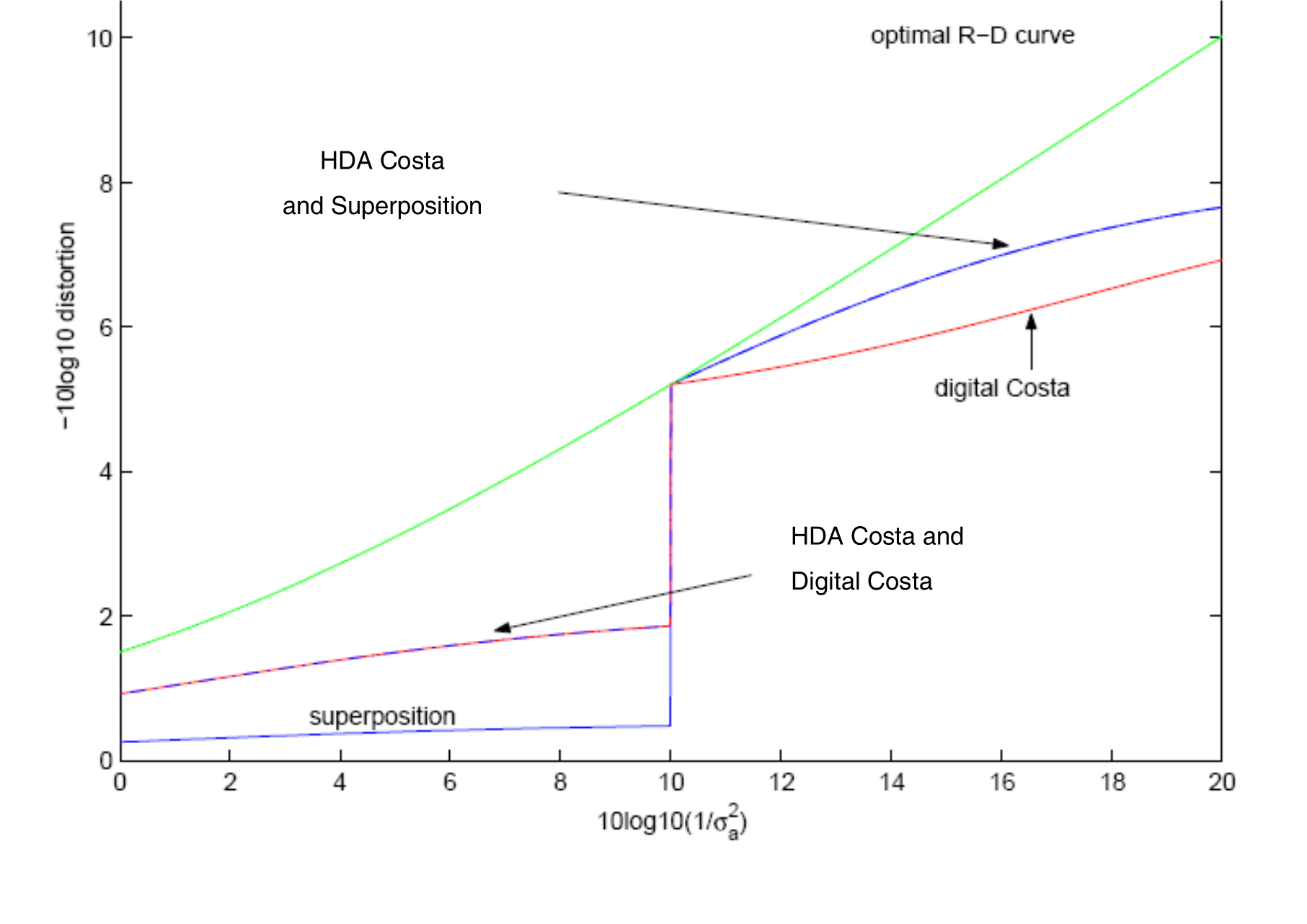}
\end{center}
\caption{Performance of different schemes for the source splitting
approach for the bandwidth compression problem with SNR mismatch.}
\label{fig:bandcompallschemes}
\end{figure}

The performance of the superposition scheme, digital Costa and HDA
Costa scheme are shown for an example with $\lambda = 0.5$ in
Fig.~\ref{fig:bandcompallschemes}. The designed SNR is defined as
$10 \log_{10} \frac{1}{\sigma^2}$ whereas the actual SNR is defined
as $10 \log_{10} \frac{1}{\sigma_a^2}$. In the example, the designed
SNR is fixed at 10dB and the actual SNR is varied from 0 dB to 20
dB.  It can be seen that the Costa coding approach is better than
superposition coding when $\sigma_a^2 > \sigma^2$ and worse for the
other case. The HDA Costa coding scheme performs the best over the
entire range of SNRs.

\section{Applications to Broadcasting with Bandwidth Compression}
\label{sec:broadcastbandwidthcomp}

We now consider the problem of transmitting $K=2N$ samples of a unit
variance Gaussian source $\vv$ in $N$ uses of the channel to two
users through AWGN channels with noise variances $\sigma_1^2$ (weak
user) and $\sigma_2^2$ (strong user) with $\sigma_1 > \sigma_2$. The
channel has the power constraint $P = 1$. We are interested in joint
source channel coding schemes that provide a good region of pairs of
distortion that are simultaneously achievable at the two users. This
problem was considered in
\cite{mittalphamdo,prabhakaran05,narayanan07}. The best known region
to date is given by the schemes therein.

\begin{figure}[htbp]
\begin{center}
\includegraphics[scale=0.35,angle = 0]{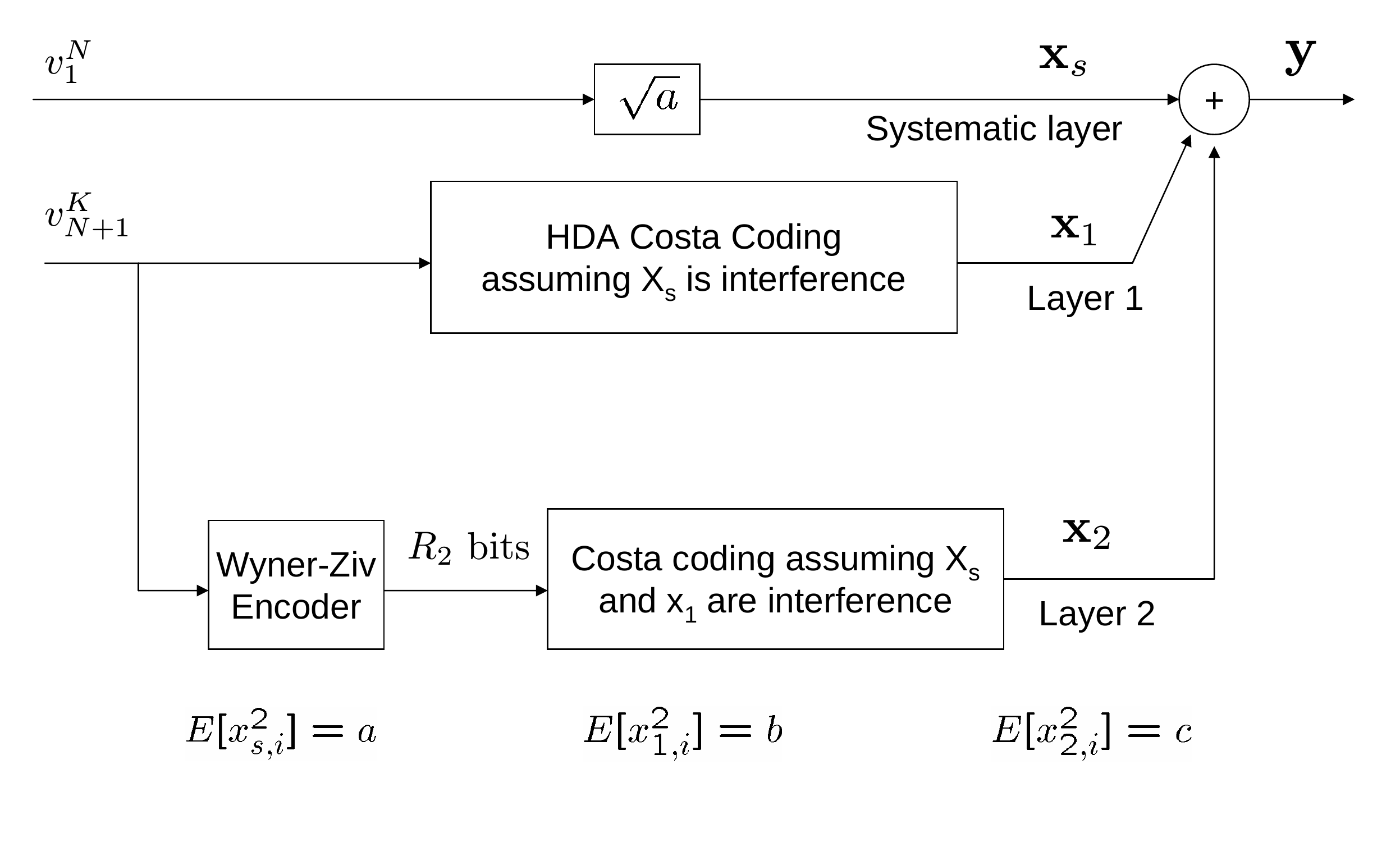}
\end{center}
\caption{Encoder model using Costa coding}
\label{fig:costasystemmodel}\end{figure}

Notice that when we design a source channel coding scheme to be
optimal for the weak user, the strong user operates under the
situation of SNR mismatch explained in
Section~\ref{sec:bandcompmismatch} with $\sigma_2^2 = \sigma^2_a < \sigma^2 = \sigma_1^2$.
Similarly, when the system is designed to be optimal for the
strong user, for the weak user $\sigma_1^2 = \sigma_a^2 > \sigma^2 = \sigma_2^2$. Motivated
by the fact that for $\lambda = 0.5$, the HDA Costa coding scheme
performs the best, we propose a scheme which is shown in
Fig.~\ref{fig:costasystemmodel}.

There are three layers in the proposed coding scheme. The first
layer is the systematic part where $N$ out of the $K$ samples of the
source are scaled by $\sqrt{a}$. Let us call this as $\xv_s =
\sqrt{a} v_1^N$. The other $K-N$ samples of the Gaussian source are
hybrid digital analog Costa coding, treating $\xv_s$ as the
interference and transmits the signal $\xv_1$ with power $b$ in the
second layer. So $\xv_1 = \uv_1 - \alpha_1 \xv_s - \kappa_c
v_{N+1}^{K}$, where $\alpha_1$ and $\kappa_c$ are the optimal
scaling coefficient to be used in the hybrid digital analog Costa
coding process and $\uv_1$ is the auxiliary variable. This layer is
meant to be decoded by the weak user and, hence, the scaling factor
$\alpha_1$ is set to be $ b/(b+c+\sigma^2_1)$. That is, this layer
sees the third layer also as {\em independent} noise.

 The third layer is first
Wyner Ziv coded  at a rate $R_2$ assuming the estimate of
$v_{N+1}^{K}$ at the receiver as side information. The Wyner-Ziv
index is then encoded using digital Costa coding assuming $\xv_s$
and $\xv_1$ are interference and uses power $c = 1 - a - b$.
Therefore, $\xv_2 = \uv_2 - \alpha_2 (\xv_s + \xv_1)$. This layer is
meant for the strong user and, hence, the scaling factor $\alpha_2 =
c/(c+\sigma^2_2)$. We then transmit $\xv = \xv_s + \xv_1 + \xv_2$.

At the receiver, from the second layer an estimate of $v_{N+1}^{K}$
is obtained. This estimate acts as side information that can be used
in refining the estimate of $v_{N+1}^{K}$ for the strong user using
the decoded Wyner-Ziv bits. The Wyner-Ziv bits are decoded from the
third layer by Costa decoding procedure.
%\begin{equation}
%R_1 = \frac 1 2 \log \left( 1 + \frac{b}{c+\sigma^2_1} \right)
%\end{equation}
%can be achieved.

The users estimate the systematic part $v_{1}^{N}$ and
non-systematic part $v_{N+1}^{K}$ by MMSE estimation from the
received $\yv$, the decoded $\uv_1$ and $\uv_2$. So the overall
distortion seen at the weak user is

\begin{equation} \nonumber D_{1} = \frac12 \frac{1}{1 +
\frac{a}{b+c+\sigma_{1}^{2}}}
  + \frac12 \frac{1}{1+\frac{b}{c + \sigma_{1}^{2}}}
\end{equation}

The distortion for the strong user is given by
\begin{equation}
\begin{split}
 & D_{2} =  \frac12 \left(
1 - [\sqrt{a} \ \alpha_{1} \sqrt{a}] \left[%
\begin{array}{cc}
  1+\sigma_2^2 & b+\alpha_{1} {a} \\
  b+\alpha_{1} a & b+\alpha_{1}^{2} a + \kappa^2\\
\end{array}%
\right]^{-1} \times \right. \\ & \left.
\left[%
\begin{array}{c}
  \sqrt{a} \\
  \alpha_{1} \sqrt{a} \\
\end{array}%
\right] \right)
  +  \frac{1/2}{1+\frac{c}{\sigma_2^2}} \left (1 - [ 0 \ \kappa] \times \right. \\ & \left. \left[%
\begin{array}{cc}
  1+\sigma_2^2 & b +\alpha_{1} {a} \\
  b+\alpha_{1} a & b + \alpha_{1}^2 a + \kappa^2\\
\end{array}%
\right]^{-1}
\left[%
\begin{array}{c}
  0 \\
  \kappa \\
\end{array}%
\right]
 \right)\\
 \end{split}
\end{equation}

The corner points of the distortion region corresponding to being
optimal for the strong and weak user respectively, can be obtained
by setting $c=0$ and $b=0$, respectively.

\begin{figure}[htbp]
\begin{center}
\includegraphics[scale=0.65,angle = 0]{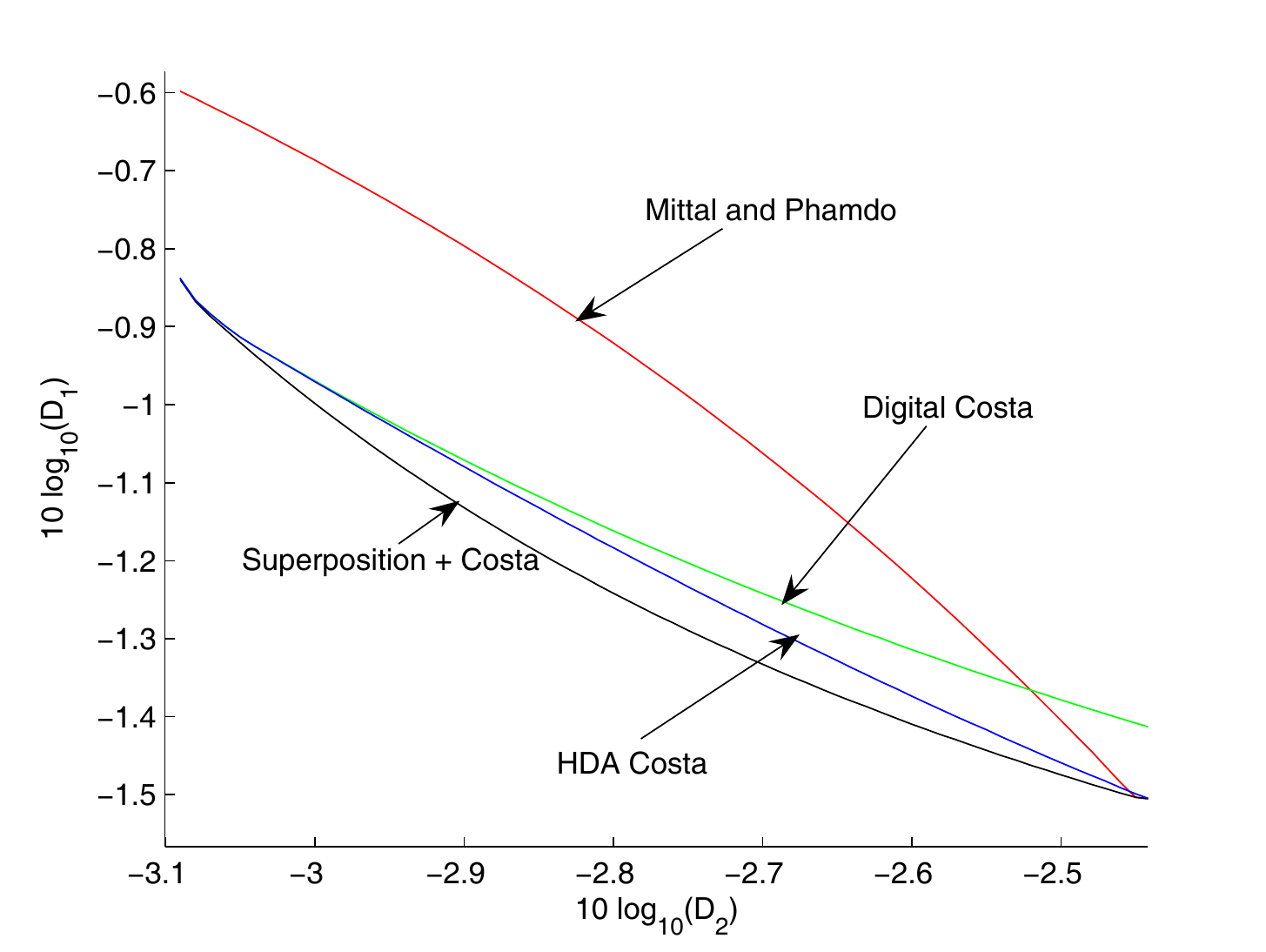}
\end{center}
\caption{Distortion regions of the different schemes for
broadcasting with bandwidth compression.}
\label{fig:broadcastresult}
\end{figure}

The distortion region for this scheme for the case of $\sigma_1^2 =
0$~dB and $\sigma_2^2 = 5$~dB is shown in
Fig.~\ref{fig:broadcastresult}. The distortion region for three
other schemes are also shown. They are the scheme proposed by Mittal
and Phamdo in \cite{mittalphamdo}, a different broadcasting scheme
which uses digital Costa coding in both the layers proposed in
\cite{narayanan07} (details can be found there) and the broadcast
scheme with one layer of superposition coding and one layer of
digital Costa coding considered in \cite{prabhakaran05,narayanan07}.
This scheme currently appears to be the best known scheme. Notice that in the
third layer, instead of using a separate Wynzer-Ziv encoder followed by
a Costa code, we could have used the HDA scheme discussed in Section~\ref{sec:combined}
with identical results.

The proposed broadcast scheme in Fig.~\ref{fig:costasystemmodel}
significantly outperforms the scheme in Mittal and Phamdo and the
digital Costa based broadcast scheme for this example. The corner
points of this scheme also coincide with those of the best known schemes
reported in \cite{prabhakaran05,narayanan07b}.

\section{Conclusion and future work} \label{sec:Conclusion}

We discussed hybrid digital analog version of Costa coding and
Wyner-Ziv coding for transmitting an analog Gaussian source
through an AWGN channel in the presence of an interferer known
only to the transmitter and side information available only to the
receiver respectively. These schemes are closely related to the
schemes by Reznic and Zamir \cite{reznic05} and \cite{kochman06},
but make the auxiliary random variable model more explicit. We
also showed that there are infinitely many schemes that are
optimal for this problem, extending the work of Bross, Lapidoth
and Tinguely \cite{shraga06} to the side information case. The HDA
coding schemes have advantages over strictly digital schemes when
there is a mismatch in the channel SNR. This makes them also
useful for broadcasting a Gaussian source to two users with
different SNRs.

\end{document}

%% file: macros.tex
\long\def\comment#1{}

\newfont{\bbb}{msbm10 scaled 700}

\newfont{\bb}{msbm10 scaled 1100}

\newcommand{\EE}{\mbox{\bb E}}

\newcommand{\ev}{{\bf e}}

\newcommand{\sv}{{\bf s}}

\newcommand{\uv}{{\bf u}}
\newcommand{\wv}{{\bf w}}
\newcommand{\vv}{{\bf v}}
\newcommand{\xv}{{\bf x}}
\newcommand{\yv}{{\bf y}}
\newcommand{\zv}{{\bf z}}